\documentclass{llncs}

\usepackage[leftcaption]{sidecap}
\usepackage{graphicx}
\usepackage{paralist}
\usepackage{geometry}
\usepackage{subcaption}
\usepackage{amsmath}
\usepackage{sidecap}
\sidecaptionvpos{figure}{c}
\usepackage{multirow}
\usepackage[english]{babel}
\usepackage{amsfonts}
\usepackage[noend]{algpseudocode}
\usepackage[open,openlevel=1]{bookmark}
\usepackage[dvipsnames,table]{xcolor}
\usepackage{hyperref}
\hypersetup{
	colorlinks   = true,    
	urlcolor     = BrickRed,    
	linkcolor    = BrickRed,    
	citecolor    =  BrickRed,      
	bookmarksopen=true
}

\everypar{\looseness=-1 }
\DeclareMathOperator*{\argmax}{arg\,max}

\usepackage[square,numbers]{natbib}
\begin{document}
\title{Community Aliveness: Discovering Interaction Decay Patterns in Online Social Communities}
\author{Mohammed Abufouda}
\institute{Department of Computer Science, University of Kaiserslautern,\\Gottlieb-Daimler-Str. 48, 67663 Kaiserslautern, Germany\\\email{abufouda@cs.uni-kl.de}}
\maketitle

\begin{abstract}
Online Social Communities (OSCs) provide a medium for connecting people, sharing news, eliciting information, and finding jobs, among others. The dynamics of the interaction among the members of OSCs is not always growth dynamics. Instead, a \textit{decay} or \textit{inactivity} dynamics often happens, which makes an OSC obsolete. Understanding the behavior and the characteristics of the members of an inactive community help to sustain the growth dynamics of these communities and, possibly, prevents them from being out of service.
In this work, we provide two prediction models for predicting the interaction decay of community members, namely: a Simple Threshold Model (STM) and a supervised machine learning classification framework. We conducted evaluation experiments for our prediction models supported by a \textit{ground truth} of decayed communities extracted from the StackExchange platform. The results of the experiments revealed that it is possible, with satisfactory prediction performance in terms of the F1-score and the accuracy, to predict the decay of the activity of the members of these communities using network-based attributes and network-exogenous attributes of the members. The upper bound of the prediction performance of the methods we used is $0.91$ and $0.83$ for the F1-score and the accuracy, respectively. These results indicate that network-based attributes are correlated with the activity of the members and that we can find decay patterns in terms of these attributes. The results also showed that the structure of the decayed communities can be used to support the alive communities by discovering inactive members.
\end{abstract}
\section{Introduction}
Nowadays, Online Social Communities (OSCs) play a vital role in our daily activities. These social networks have become the main source for sharing news, connecting people, communicating, eliciting information, and finding jobs. Thus, studying the behavior and the dynamics of the members of these online platforms is crucial for sustaining these communities and maintaining the services they provide. One way to model these communities is \textit{network} representation, where nodes represent the members of the social platform and edges represent the interactions between the members. Since the seminal works by Barab{\'a}si and R{\'e}ka~\cite{barabasi1999} and by Watts and Strogatz~\cite{watts1998collective}, the field of \textit{network science} has witnessed a huge amount of research into the dynamics of systems represented as networks. Social networks have been, and are being, studied as an example of networks that contain a lot of dynamics overtime. While a lot of social networks have been successful in sustaining their aliveness and growth dynamics, many others have experienced decay dynamics. Online social platforms such as MySpace and Friendster are now out of service due to the huge decay they have encountered, causing a massive drop of their market values. 

In this work, we are interested in understanding the decay dynamics of OSCs and the interaction patterns that accompany, or possibly cause, community decay. Gaining insights into the decay interaction patters among members of OSCs will enable us to better understand the decay process, and hence, help to provide possible actions by prolonging the life of these communities and supporting their resilience to inactivity disruptions.
More precisely, we predict members who will leave a network (or become inactive) by using a Simple Threshold Model (STM) and by using a machine learning supervised binary classification framework employing network-based and exogenous features\footnote{Exogenous means any information that is not based on the structure of a network.}. The contributions of this paper can be described as follows: (1) an exploratory data analysis of the decayed StackExchange communities and a comparison with the ones that are alive supported with a \textit{ground truth}; (2) a Simple Threshold Model (STM) for predicting social inactivity using network-based measures or members' exogenous information; (3) a machine learning framework for predicting social leave using network-based measures and members' exogenous information; (4) guidelines for feature selection in predicting member's inactivity.
The results provided insights regarding the network-based properties and also the exogenous features of inactive members that are correlated with social inactivity. These insights may help to prevent decay dynamics, to engineer resilient social networks, and to express the aliveness of OSCs.
\\The reminder of this paper is structured as follows. Section~\ref{sec:relatedwork} provides a survey of related work. Section~\ref{sec:prelimiarydefinitions} gives the required definitions used in this paper. In Section~\ref{sec:datasetandEDA}, we describe the datasets used, provide the preliminary results of the exploratory data analysis we conducted, and formulate the research questions. The methods we used are described in detail in Section~\ref{sec:method}. The results are presented in Section~\ref{sec:results}. In Section~\ref{sec:discussion} the research questions are answered and the results are discussed. This paper concludes in Section~\ref{sec:conclusion}, which also provides an outlook regarding the future directions.
\section{Related Work}
\label{sec:relatedwork}
Dorogovtsev and Mendes~\cite{dorogovtsev2000} studied mathematically the decay properties of the networks and found similar characteristics of the preferential attachment provided earlier by Barab{\'a}si and R{\'e}ka~\cite{barabasi1999}. Newman et al.~\cite{jin2001} studied the growth dynamics of social networks between mutual friends and provided a model that showed similar characteristics as of these real networks. The growth dynamics was then studied extensively in many researches for different domains. For example, Newman~\cite{newman2001} studied the growth dynamics, namely the clustering and preferential attachment, of scientific collaboration networks in physics and biology fields. Similarly, Bornholdt et al.~\cite{ebel2002dynamics} provided another model for simulating the growth dynamics of social network. With the availability of datasets, Barab{\'a}si et al.~\cite{barabasi2002evolution} provided an empirical study on the evolution of the collaboration patterns of the scientific collaboration networks. Leskovec et al.~\cite{leskovec2005} studied the growth dynamics of networks by observing some repeated patterns, namely densification laws and shrinking diameters. Backstrom et al.~\cite{backstrom2006} investigated the growth dynamics of group formation and community memberships in online social networks. They provided a model for predicting when a member would join a community in a social network. A preferential attachment growth model was presented by Capocci et al.~\cite{capocci2006} to study the growth dynamics of the Wikipedia online encyclopedia. Similarly, Kossinets and Watts~\cite{kossinets2006} studied the growth dynamics of a social network of students, faculty, and staff members of a university. They found that the evolution of the network was mainly affected by the network structure itself and some other external organizational structure. Kumar et al.~\cite{Kumar2006} provided a large scale analysis of the social network evolution on five million members and more than ten million relationships. Their analysis revealed some structural properties of the growth process in online social networks. Ahn et al.~\cite{ahn2007} studied the growth of Myspace and Orkut, before they were permanently closed, as a real examples of networks with growth dynamics. They studied the scaling behavior of the degree distribution over time for these networks and found that they have different exponents. Mislove et al.~\cite{mislove2008} studied extensively the growth of the Flicker online social network and found link formation patterns.\\
The aforementioned works concentrated the \textit{growth dynamics}. However, the dynamics of social network is not limited to growth dynamics, but also includes \textit{decay} dynamics that may occur in the social network which leads to inactive (decayed) social network. Social network platforms like Orkut, MySpace, Friendster, and Friendfeed are now out of service after being active and growing for long time. There are few research that addressed the problem of inactivity in social networks. For example, Garcia et al~\cite{Garcia2013} studied the properties of different networks (decayed and active ones) in terms of $k$-core analysis. Later, Malliaros and Vazirgiannis~\cite{malliaros2013} provided a method to quantify and measure the \textit{engagement} of the members. Their measures enabled assessing the robustness of the networks over time. In a related vein, Wu et al.~\cite{Wu2013} provided a method for understanding the dynamics of social engagement of the members of the co-authorship social network of the DBLP. They showed that there was a correlation between actions of the departed members in the studied datasets. They also provided some insights regarding the properties of the members who departed the networks. Cannarella and Spechler provided an epidemic model for predicting the dynamics of the members of the Facebook~\cite{cannarella2014}. The results showed that the Facebook would lose $80\%$ of its users between $2015$ and $2017$, which did not happen until now. Karnstedt et al.~\cite{karnstedt2011}, Kawale et al.~\cite{Kawale2009}, and Wang et al.~\cite{Wang2016} in a recent work provided prediction models for users lifespan in online social settings, also called \textit{users churn}.\\
Our work is different from the previous works from two perspectives. Firstly, users churn normally has a one-to-many relationship between members and service provider. In this scenario, the social interaction, which is our main concern, is very limited. In this work, the social interaction between the users is the main concentration in the models that we provide. Secondly, our main concern is the decay of the social interaction between humans in online social networks in order to better understand the decay dynamics in online social networks\footnote{More information about our view of social decay can be found in our previous work~\cite{abufouda2016,abufouda2017}.}.
\section{Preliminary definitions}
\label{sec:prelimiarydefinitions}
\subsection{Networks and measures}
\label{subsec:networks and measures}
An undirected graph $G=(V,E)$ is defined as a tuple of two sets $V$ and $E$, such that $V$ is the set of nodes and the set $E$ is the set of edges. An edge $e$ is defined as $e=\{u,v\}$ where $u,v \in V$ and $e \in E$.
\begin{table}[]
\centering
\footnotesize
\begin{tabular}{|l|p{11cm}|}
\hline
Measure            & Description                                                                                                                                                                                                                                                                                                                                  \\ \hline
D(v)               & The \textit{Degree} of a node $v$, $D(v)= |\Gamma(v)|$, is the cardinality of the set $\Gamma(v)$.                                                                                                                                                                                                                                         \\ \hline
B(v)               & The \textit{Betweenness} of a node $v$ is defined as: $B(v)=\sum_{s\in V(G)}\sum_{t\in V(G)}\frac{\sigma_{st}(v)}{\sigma_{st}}$, where $\sigma_{st}(v)$ is the number of shortest paths between the nodes $s$ and $t$ that includes the node $v$ and $\sigma_{st}$ is the number all shortest paths between the nodes $s$ and $t$.         \\ \hline
$\mathcal{C}(v)$   & The \textit{Closeness} of a node $v$ is defined as: $\mathcal{C}(v)=(\sum_{w\in V(G)}d(v,w))^{-1}$, where $d(v,w)$ is the distance between the nodes $v,w$.                                                                                                                                                                                \\ \hline
$Core(v)$          & A $k$-core subgraph of a graph $G$ is the maximal subgraph such that each node has a degree at least $k$. The \textit{coreness} of a node $Core(v)=k$ if the node $v$ is in the $k$-core subgraph and not in the $k+1$-core subgraph.                                                                                                      \\ \hline
$E(v)$             & The \textit{Eccentricity} of a node $v$, $E(v)$, is the maximum distance between the node $v$ and a node $u$.                                                                                                                                                                                                                             \\ \hline
Node-Cut & A $node$-$cut$, sometimes called articulation point, in a connected graph is a node whose removal increases the number of components in the graph.                                                                                                                                                                                          \\ \hline
$\mathcal{MC}(v)$  & A \textit{minimum cut} of two nodes $u,v$, $MinCut(u,v)$ is the minimum number of edges that are required to be removed in order to separate the two nodes. The averaged minimum cut of a node $v$ is defined as: $\mathcal{MC}(v)=\frac{1}{n} \sum_{u \in E, u \neq v}{MinCut(u,v)}$, where $n$ is the number of nodes in a graph. \\ \hline
\end{tabular}
\caption{The definitions of the used network-based measures.}
\label{tab:networkmeasures}
\end{table}
The set of neighbors of a node, $\Gamma(v)$, is the set of nodes that are connected to the node $v$. Table~\ref{tab:networkmeasures} shows a list of network-based measures.
\subsection{Binary classification}
\label{subsec:binaryclassification}
Given a binary variable $Y=\{$\textit{True}, \textit{False}$\}$, and a set of attributes $X=(x_1,x_2,\cdots,x_n)$ that are assumed to affect the value of $Y$. Then, a probabilistic supervised binary classifier in its simple forms is defined as $P(Y$=True$| X)=[0,1]$ and a threshold to binarize the probability result. There are many binary classifiers and optimization techniques for finding the best model during the learning process. Describing these classifiers and learning process is beyond the goal of this paper. 
\begin{table}[]
\centering
\footnotesize
\begin{tabular}{|l|p{11cm}|}
\hline
Measure       & Description                                                                                                                                                                                                                                                                                  \\ \hline
$\mathcal{P}$ & The \textit{Precision} is defined as: $\mathcal{P} = \frac{TP}{TP+FP}$.                                                                                                                                                                                                                    \\ \hline
$\mathcal{R}$ & The \textit{Recall} is defined as: $\mathcal{R} = \frac{TP}{TP+FN}$.                                                                                                                                                                                                                       \\ \hline
$\mathcal{A}$ & The \textit{Accuracy} is defined as: $\mathcal{A} = \frac{TP+TN}{TP+FN+TN+FP}$.                                                                                                                                                                                                            \\ \hline
$F1$-$score$  & The harmonic mean of the precision and the recall, the F1-score, is defined as $F= 2\cdot\frac{\mathcal{P}\cdot \mathcal{R}}{\mathcal{P}+\mathcal{R}}$.\\ \hline
\end{tabular}
\caption{Prediction performance measures.}
\label{tab:predictionmeasures}
\end{table}
To evaluate the efficiency of a classifier, we have different measures. A true-positive (TP) happens when binary classifier classifies an instance as \textit{True} where its real value is \textit{True}. Similarly, we have true-negative (TN), false-positive (FP), and false-negative (FN). Based on those simple metrics, Table~\ref{tab:predictionmeasures} shows the used prediction performance measures.
\section{Dataset and research questions}
\label{sec:datasetandEDA}
\subsection{The StackExchange dataset}
\label{subsec:theStackExchangedataset}
\begin{figure}[!ht]
    \centering
    \begin{subfigure}[b]{0.48\textwidth}
    	\centering
        \includegraphics[width=\linewidth]{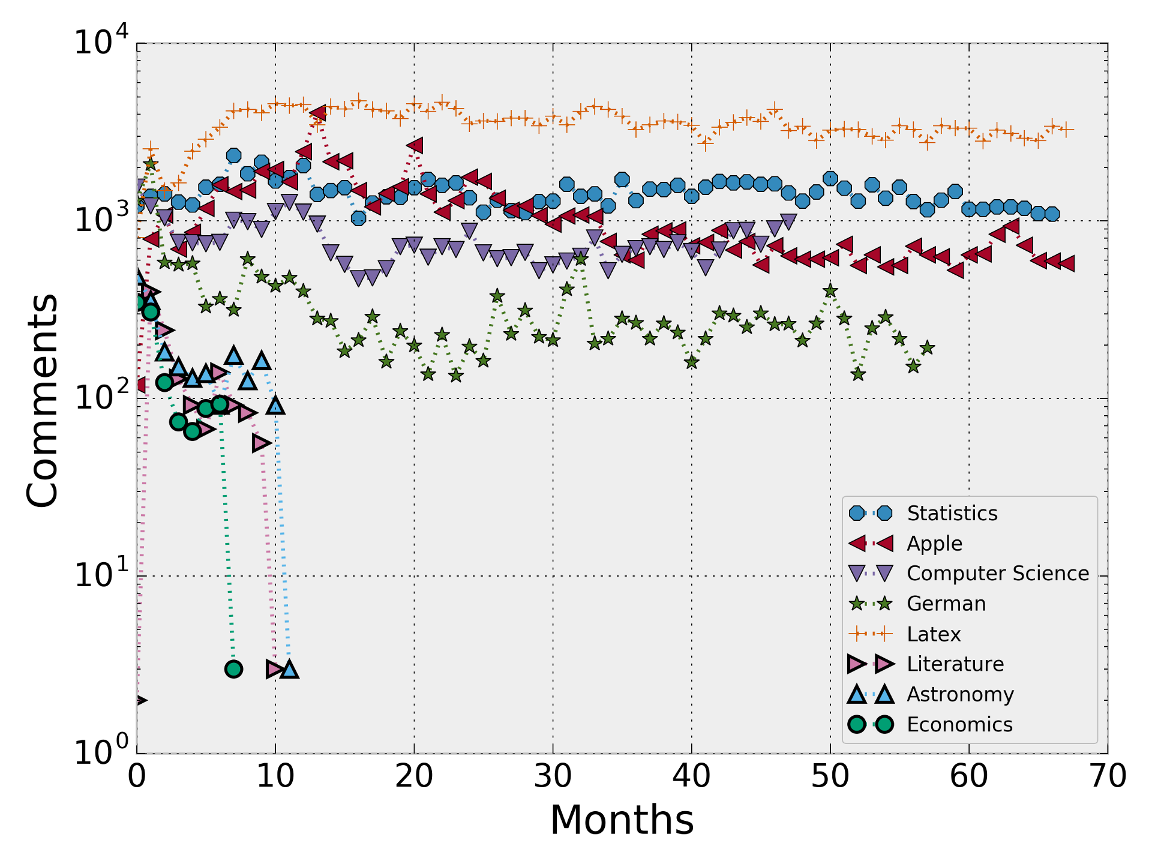}
        \caption{Comments}
    \end{subfigure}
    \begin{subfigure}[b]{0.48\textwidth}
        	\centering
        \includegraphics[width=\linewidth]{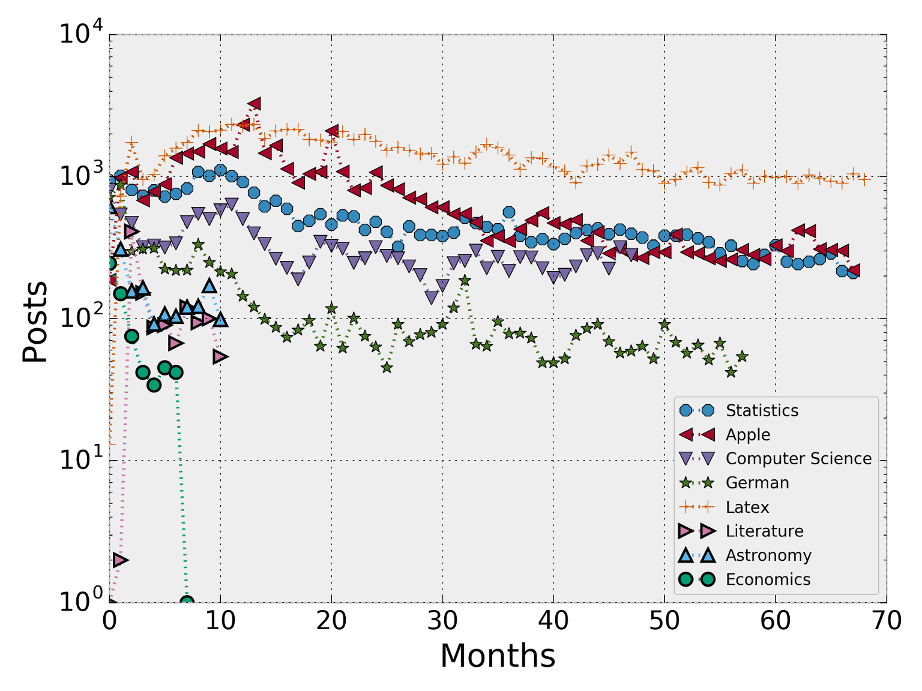}
        \caption{Posts}
    \end{subfigure}
     \caption{(Color online) The activity of members of some communities of the StackExchange websites in terms of \textit{comments} and \textit{posts} count over time. The x-axis represents the number of months since the launch of a website. Communities with bold markers, \textit{Literature}, \textit{Astronomy}, and \textit{Economics}, were closed after the failure of their beta versions.}
        \label{fig:commentsandposts}
\end{figure}

\begin{figure}
  \centering
    \includegraphics[scale=1]{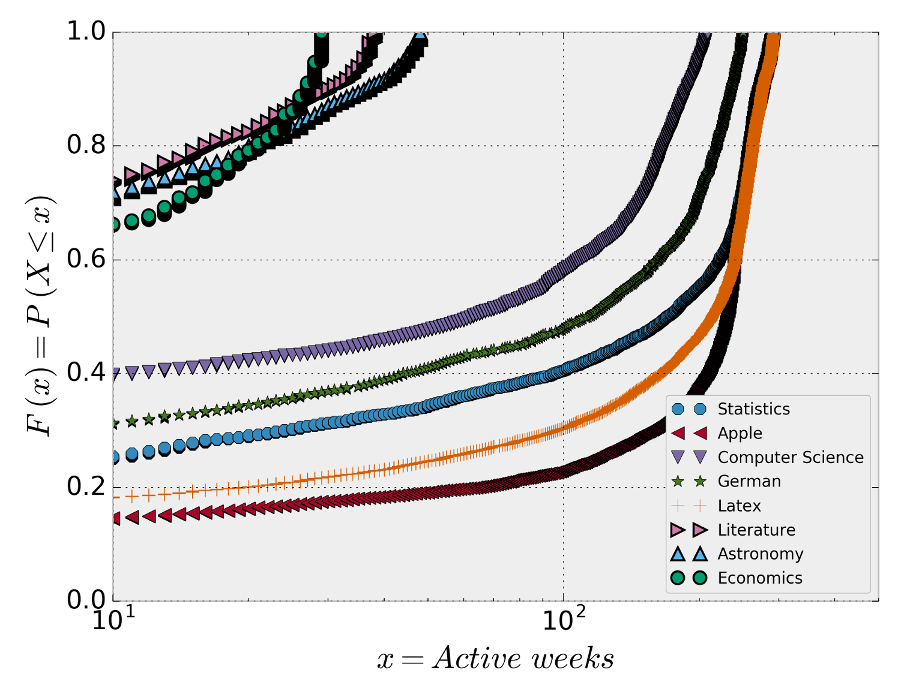}
  \caption{(Color online) The figure shows the Cumulative distribution function (CDF) of members' active weeks. The number of active weeks is calculated as the difference between the last log-in date and the registration date of a member. The CDF is then calculated as $F(x) = P(X\leq x)$. Note that the x-axis is log-scaled. }
  \label{fig:activeperiod}
\end{figure}

\begin{figure*}[!t]
	\centering
	\parbox{3cm}{
		\includegraphics[width=3cm]{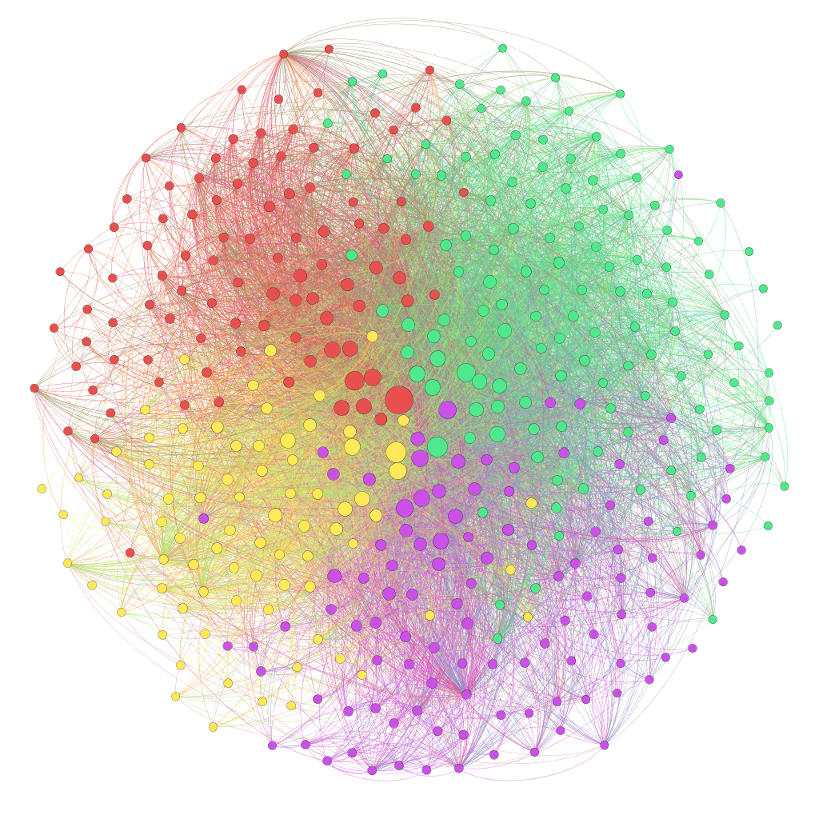}
		\caption*{Oct-2009}
		\label{fig:toy1}}
	\qquad
	\begin{minipage}{3cm}
		\includegraphics[width=3cm]{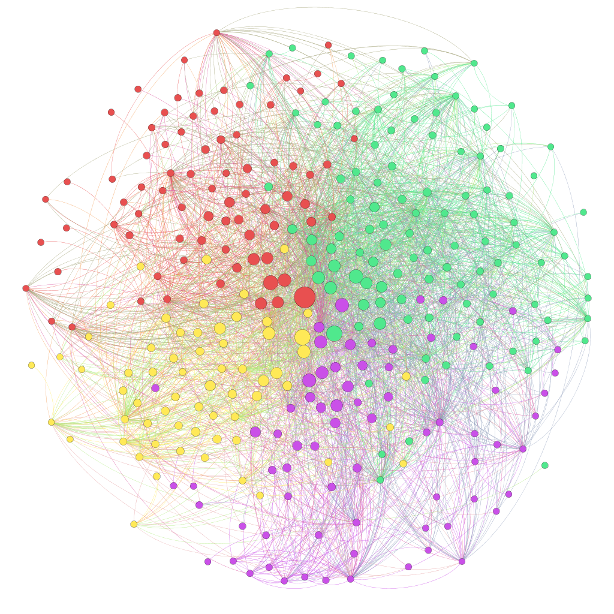}
		\caption*{Feb-2010}
		\label{fig:toy3}
	\end{minipage}
	\qquad
	\begin{minipage}{3cm}
		\includegraphics[width=3cm]{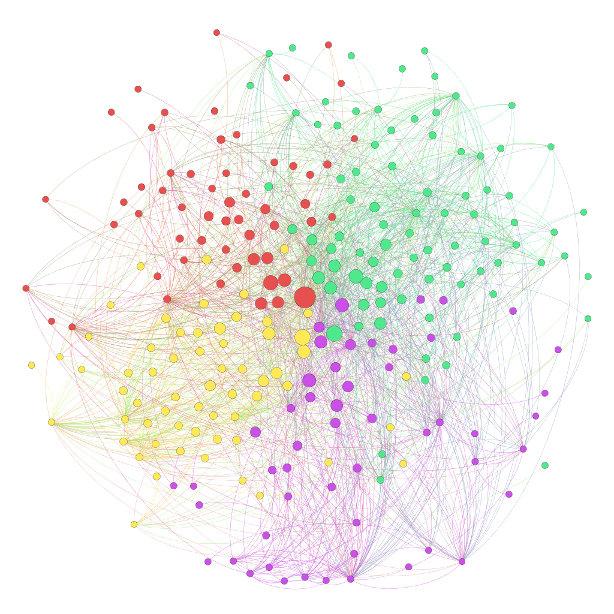}
		\caption*{Apr-2010}
		\label{fig:toy4}
	\end{minipage}
	\qquad
	\begin{minipage}{3cm}
		\includegraphics[width=3cm]{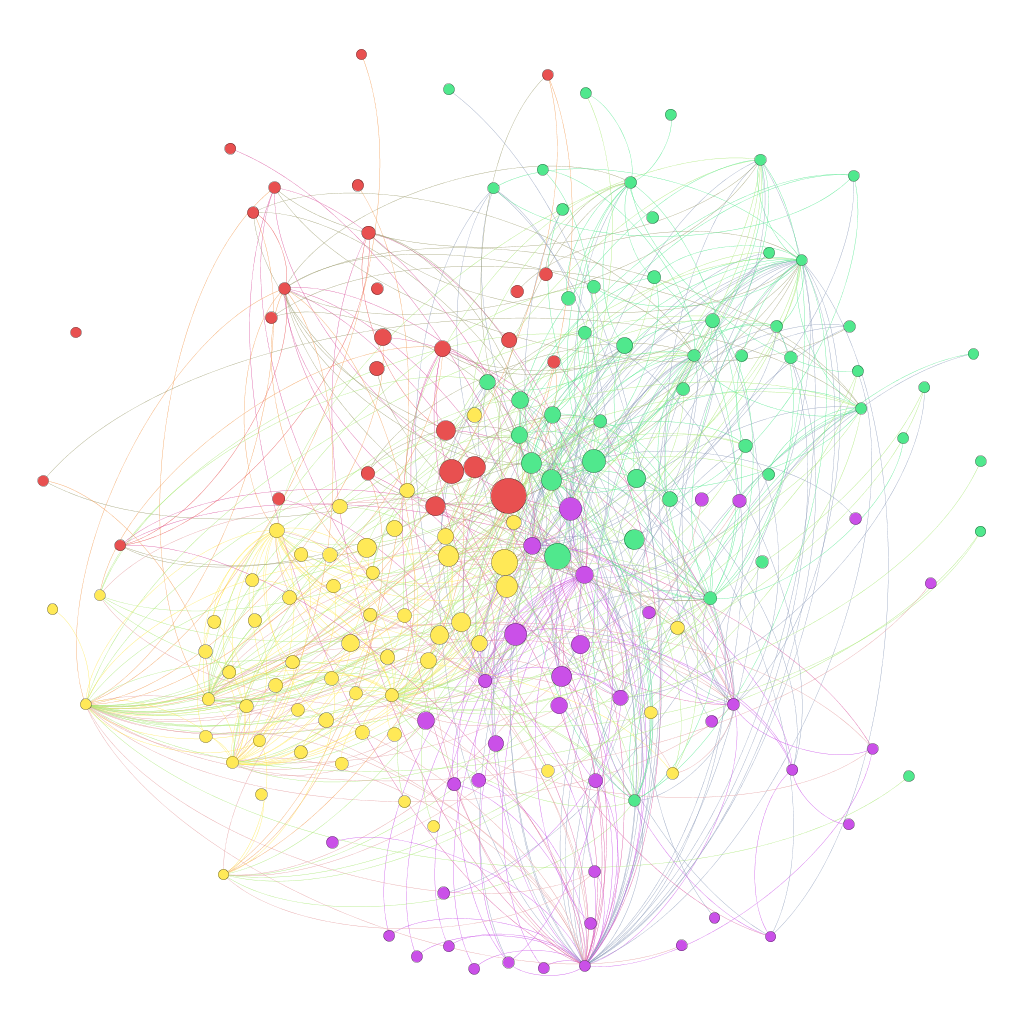}
		\caption*{Aug-2010}
		\label{fig:toy6}
	\end{minipage}
	\qquad
	\begin{minipage}{3cm}
		\includegraphics[width=3cm]{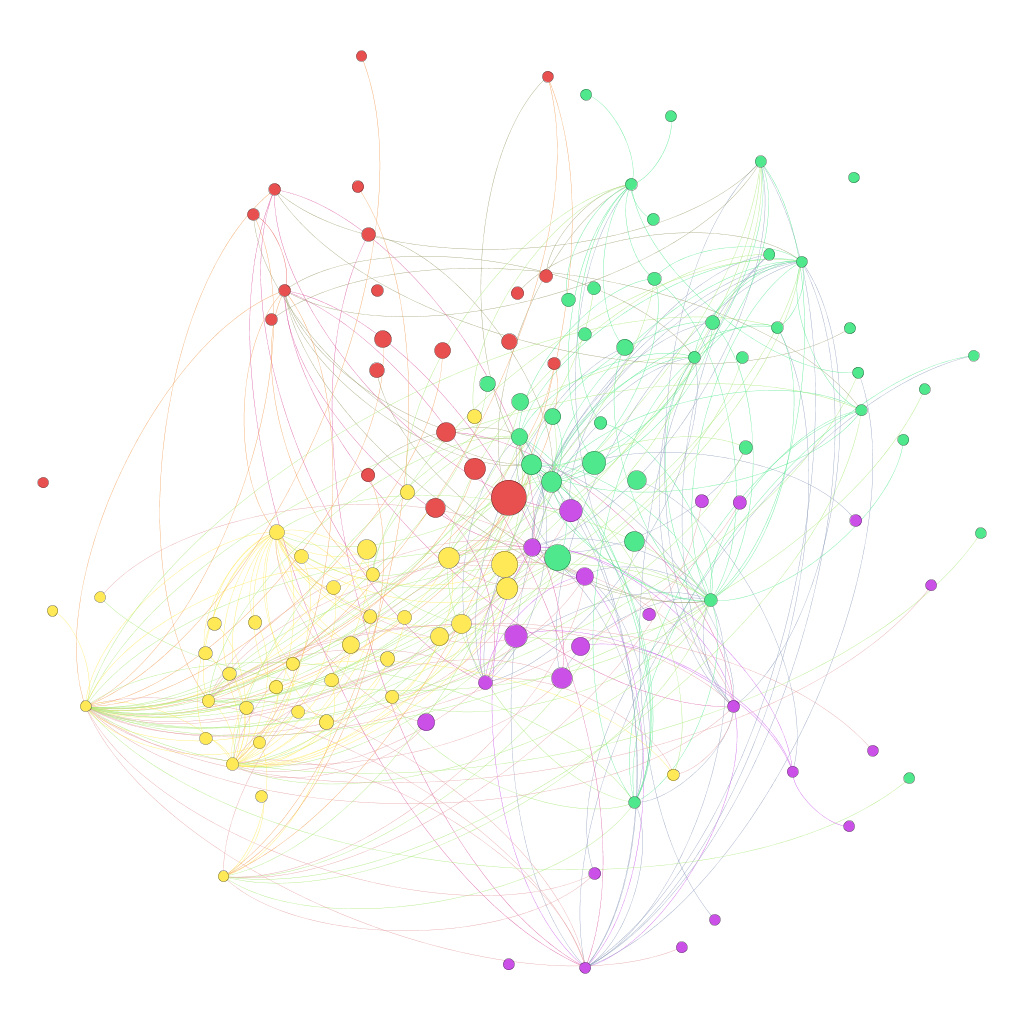}
		\caption*{Oct-2010}
		\label{fig:toy7}
	\end{minipage}
	\qquad
	\begin{minipage}{3cm}
		\includegraphics[width=3cm]{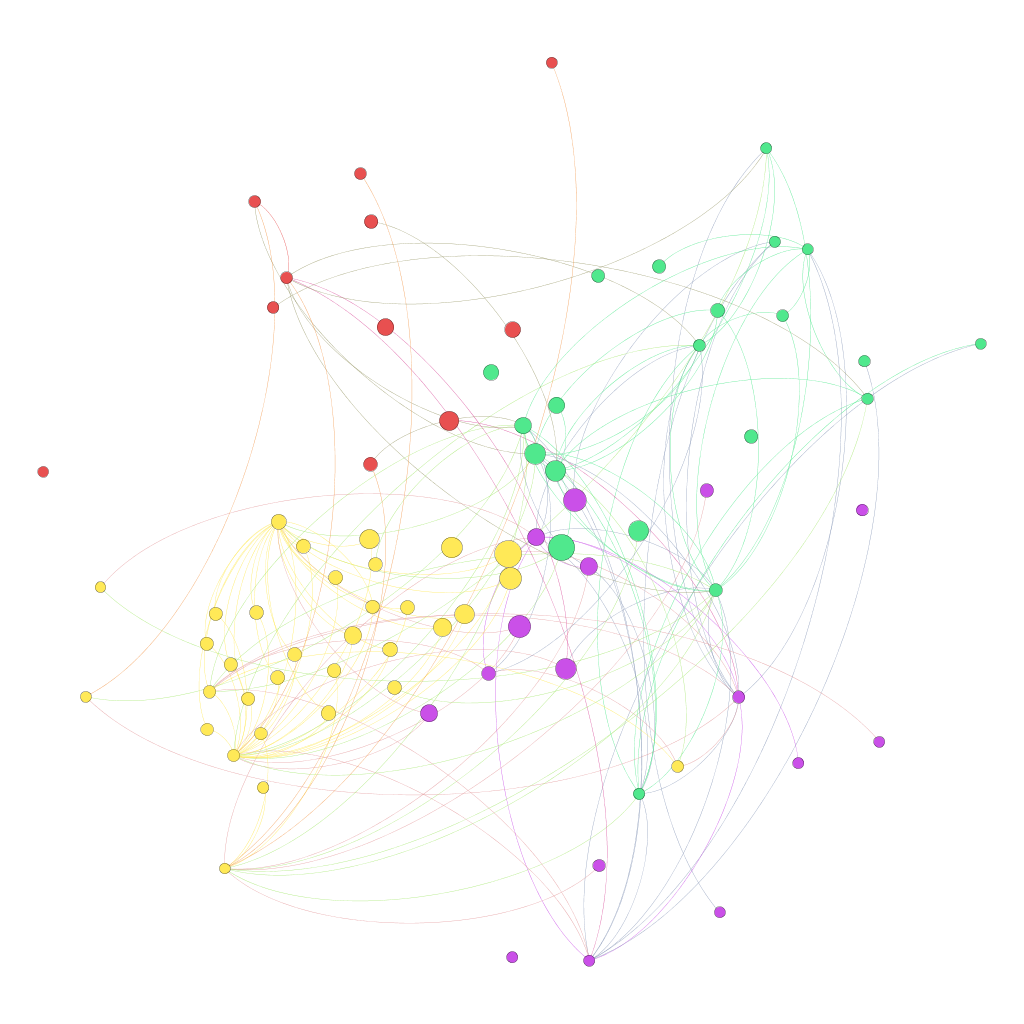}
		\caption*{Jan-2011}
		\label{fig:toy8}
	\end{minipage}
	\qquad
	\caption{(Color online) The network representation of real data of the \textit{Business Startups} website social network between Oct-2009 and Jan-2011. The networks (nodes are the members and edges represent interaction among them) were colored by clusters and having a node size directly proportional to its degree. For the sake of simplicity and for a better visualization, we restricted the networks to core members who registered-in during the first four months of the website. The networks in the figure show a clear decay of the number of nodes and edges.}
	\label{fig:networksDecay}
\end{figure*}

The StackExchange\footnote{https://StackExchange.com/} is a portal that includes many question \& answer websites for different focused topics. Any of these websites firstly starts as a beta community until it shows potential for permanent public access. However, not all of these beta communities succeed in providing the required activity and expert attraction to sustain a growth. Thus, these beta communities close. There are many examples of closed StackExchange beta communities\footnote{A list of the closed websites can be found here: http://bit.ly/2bVeukz} and the content generated by the users during the beta versions of them is still available. We downloaded, parsed, structurized, and analyzed a list of the closed communities in order to understand what is going during the decay (inactivity) dynamics in social networks\footnote{The data, the code, and the related material of this dataset are available for the public upon request.}. Figure~\ref{fig:commentsandposts} shows the activity, in terms of the number of posts and the number of comments, of different websites of the StackExchange. The figure shows that the activity of the members is almost stable in the alive communities, \textit{Statistics}, \textit{Apple}, \textit{Computer Science}, \textit{German}, and \textit{Latex} while the activity of the members is decaying in the closed communities, \textit{Economics} \textit{Literature} and the \textit{Astronomy}.
Figure~\ref{fig:activeperiod} shows the difference between the decayed and the alive communities in terms of the active weeks. The decayed communities exhibit a significantly shorter active weeks compared to alive communities. Figure~\ref{fig:networksDecay} shows the network representation of real data of the \textit{Business Startups} website social network. This website was closed after a decay in its activity and we used it as a ground truth data in our experiments because it has the longest beta time which enabled us to have more meaningful snapshots overtime. The ground truth in the settings of this research is defined as the following.
An observed (real) network and a predicted network are defined as $G=(V,E)$ and $G'=(V',E')$, respectively. As we are interested in predicting the nodes, we define false-negative nodes as the nodes in the set $V \setminus V'$, i.e., the set of nodes that exist in the observed network and the prediction model missed their existence. Likewise, the set $V' \setminus V$ is the set of false-positive nodes, i.e., the set of nodes that the prediction model predicted while they are not present in the observed data. Additionally, the set $V \cap V'$ is the true-positive nodes. Note that there is no true-negative as we predict only the \textit{initial} nodes (cf. Figure \ref{fig:toy}).

\subsection{Experiment setup and research questions}
\label{subsec:experimentsetupresearchquestions}
\begin{figure}[!ht]
	\centering
	\begin{subfigure}[b]{0.3\textwidth}
		\centering
		\includegraphics[width=\linewidth]{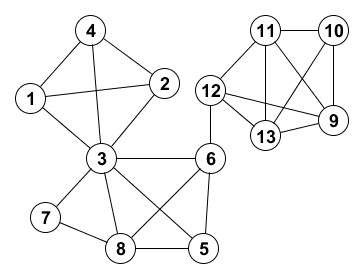}
		\caption{$G_t$}
		\label{fig:toy_1}
	\end{subfigure}
	\begin{subfigure}[b]{0.3\textwidth}
		\centering
		\includegraphics[width=\linewidth]{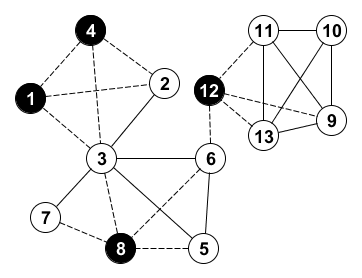}
		\caption{$G_{t_1}$}
		\label{fig:toy_2}
	\end{subfigure}
	\begin{subfigure}[b]{0.3\textwidth}    
		\centering
		\includegraphics[width=\linewidth]{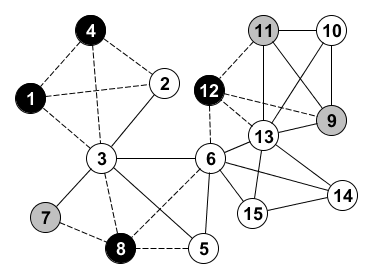}
		\caption{$G_{t'}$}
		\label{fig:toy_3}
	\end{subfigure}
	\caption{A schematic illustration shows the different networks, $G_t$, $G_1$, and $G_{t'}$, over time where $t < t_1 < t'$. The nodes in network $G_t$ are \textit{initial} nodes. The training period is performed on the time $t$ to $t_1$, where the black nodes in the network $G_{t_1}$ are the observed nodes that left the network. Then, we test on the network $G_{t'}$ where we predict the inactivity of other nodes, e.g., the grey nodes in $G_{t'}$. Note that nodes that emerge in the network $G_{t'}$ and are not in the initial node set, e.g., nodes 14 and 15, are ignored.}
	\label{fig:toy}
\end{figure}

We define the \textit{Members leave} problem as follows. Given a network $G_t=(V_t,E_t)$ that represents the network at time point $t$, and another version of the network, $G_{t'}=(V_{t'},E_{t'})$, where $t'>t$. Then, we predict the set of nodes $V_{t'}$. During a training phase, we observe the networks $G_t$ and $G_{t_1}$, then we model the properties of the nodes $V_t \setminus V_{t_1}$ in order to predict the set of nodes $V_t \setminus V_{t'}$. The set of nodes $V_t$ are called the \textit{initial} nodes, and any node $u$ such that $u \in V_{t'}$ and $ u \notin V_{t}$ is ignored for all $t < t_1 < t'$. The \textit{inactive} members (members who left) are the set of nodes $V_t \setminus V_{t'}$. Figure~\ref{fig:toy} shows an example of the training and testing phases on exemplar networks.
Based on that, we formulate our research questions as follows:\\
\textbf{RQ1}:\textit{ How efficient is it to predict members leaving a social community using network-based measures?} By answering this question we aim at understanding how efficient is it to use the network topology in understanding networks decay dynamics.\\
\textbf{RQ2}: \textit{What are the network-based properties for the members who left or about to leave a community?} By answering this question we want to get insights regarding the properties of the nodes before leaving the network. Thus, networks and community maintainers can provide counter actions when a decay process starts to emerge, which enables sustaining resilient networks.\\
\textbf{RQ3}: \textit{How helpful are the exogenous attributes in predicting members leaving?} Obtaining additional information other than the network representation is not always possible. Thus, answering this question will give us more insights on whether the network-based attributes contains sufficient information to predict the members leave. Also, we compare the prediction results performed using only the network-based attributes or only the exogenous attributes.\\
\textbf{RQ4}: \textit{Do decayed communities embrace leave patterns that can be used to study the inactivity of communities that are alive?} The alive communities of the StackExchange may suffer from members inactivity, however, this inactivity is mitigated by the activity of new members and new discussions that support the aliveness of these communities and make them active until today. Answering this question will give us insights regarding whether there are community-independent decay patterns or not that can be used to track potential decay of the alive communities.

\section{Method}
\label{sec:method}
In this section, we describe our method which contains the feature model that we built and used in the prediction in addition to two models for predicting \textit{members leave}.
\subsection{Features model}
\label{subsec:featuremodel}
We provide two types of features for the nodes of studied communities as follows:\\
\textbf{(1)} \textit{Network-based measures}: which are the values of node's attributes that are based on network measures presented in section~\ref{subsec:networks and measures}. These measures reflect how a node is connected in the network. For each node $v \in G_t$ we calculate a set $\vartheta_v$ that represents node's network-based attribute values.\\
\textbf{(2)} \textit{Exogenous attributes}: which are the values of node's non-network attributes of the members. For each $v \in G_t$ we obtain the set of attribute value $\theta_v$. For the StackExchange dataset, these attributes include the \textit{Upvotes} and the \textit{Downvotes} a member received, the profile \textit{View} counts, and the \textit{Reputation}, among others.\\
In addition to the above two types, we have \textit{Leave label}, which indicates whether a node $v \in G_t$ left the network at time $t_1$ or not.
Thus, an instance of the feature model for a node $v \in G_t$ is defined as $I_v= \vartheta_v \cup \theta_v \cup \{True\mid False\}$. The resulted feature model is then defined as $\mathcal{M}(G_t,G_{t_1})=\{I_{v}, \forall v \in G_t \}$ and used during the training phase to predict the node leave at $t'$, where $t<t_1<t'$.

\subsection{Simple Threshold Model (STM)}
\label{subsec:simplethresholdmodel}
\begin{figure*}[!ht]
\centering
\includegraphics[scale=0.85]{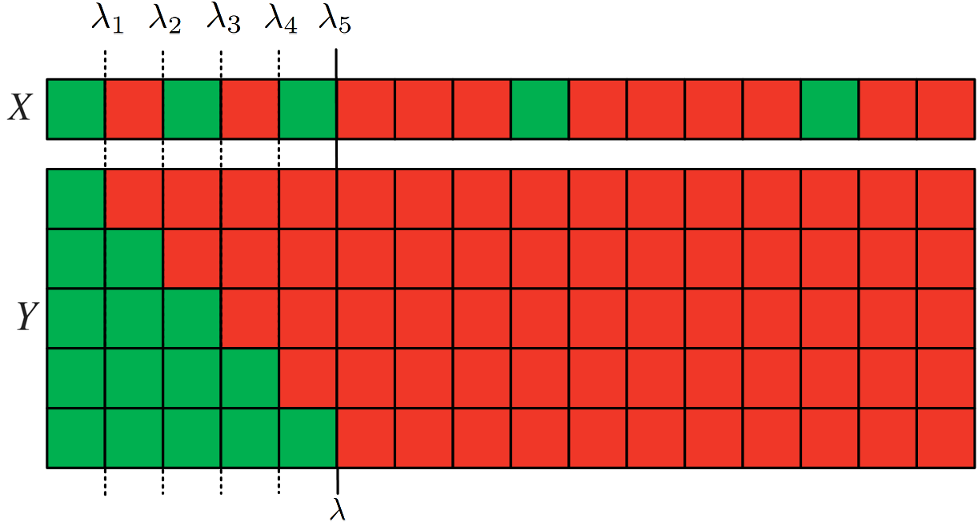}
\caption{(Color online) The diagram shows an example on how the value of $\lambda$ is computed during the training phase. The set $X$ represents a sorted vector of the values of one attribute, say the \textit{Betweenness} of a node, of $\mathcal{M}(G_t,G_{t_1})$ excluding the leave label for the network $G_t$. For the vector $X$, green cells mean that the node left at $t_1$ and the red cells mean that the node did not leave at $t_1$. The goal is to find a vector $Y$ that is composed of two vectors each of them has the same value for all of its elements: True or False (in the figure green or red). The model aims at finding the best value $\lambda$ of the \textit{Betweenness} such that it maximizes the prediction performance, e.g., the F1-score. That is, the $X$ vector is the actual labels and the $Y$ vector is the predicted labels. The chosen $\lambda = \lambda_5$ is because the values of the F1-score are $ 0.3,0.29,0.5,0.4,$ and $0.6$ for $\lambda_1$,$\lambda_2$,$\lambda_3$,$\lambda_4$, and $\lambda_5$, respectively. 
Later values for $\lambda$, i.e., $\lambda_q$ for $q>5$, have F1-scores less than $0.6$.}
\label{fig:threshold}
\end{figure*}

We present a simple model for predicting users leave using only one attribute $i \in \vartheta_v \cup \theta_v$. The idea is that, for this attribute $i$ we find its value for all nodes and sort them. Afterward, we find the best threshold value $\lambda$, that splits the nodes into two disjoint sets such that each element in each set has the same label, \textit{True} or \textit{False}. The threshold value is chosen such that it maximizes one of the prediction measures provided in section~\ref{subsec:binaryclassification}. Figure~\ref{fig:threshold} shows a schematic diagram of this model. More formal, 
given sorted array of the values of an attribute $i$ defined as $ values(i) $ such that $ i \in \vartheta \cup \theta $, and the corresponding leave label array. Let $f$ be a function defined as $f: \lambda \rightarrow s$, where $s$ is one of the prediction metrics, then the STM is defined as:
\begin{equation*}
\argmax\limits_{\lambda}{f(\lambda) = \{\lambda \ | \ \lambda \in values(i)\} }.
\end{equation*}
For binary attributes, like the \textit{articulation points}, we do not need to find the threshold $\lambda$, instead we calculate the prediction performance measure directly on the sets that contain the values of articulation points (i.e., predicted values) and the leave label (i.e., actual values).
\subsection{Machine learning classification}
\label{subsec:machinelearningclassification}
With the STM, we can only benefit from the information provided by one attribute at a time. To incorporate more attributes, we used the whole feature model $\mathcal{M}$ for training and testing a supervised machine learning binary classifier to predict the \textit{leave label}. For the evaluation, we used the same evaluation metrics presented in Section~\ref{subsec:binaryclassification}. We used the Support vector machines, the Logistic regression, and the Random Forests classifiers' implementation of the $scikit$-$learn$~\cite{scikitlearn}.
\section{Results}
\label{sec:results}
\subsection{Prediction using one attribute}
\label{subsec:predictionusingoneattribute}
\begin{figure*}[!ht]
\centering
\parbox{3.5cm}{
\includegraphics[width=4.2cm]{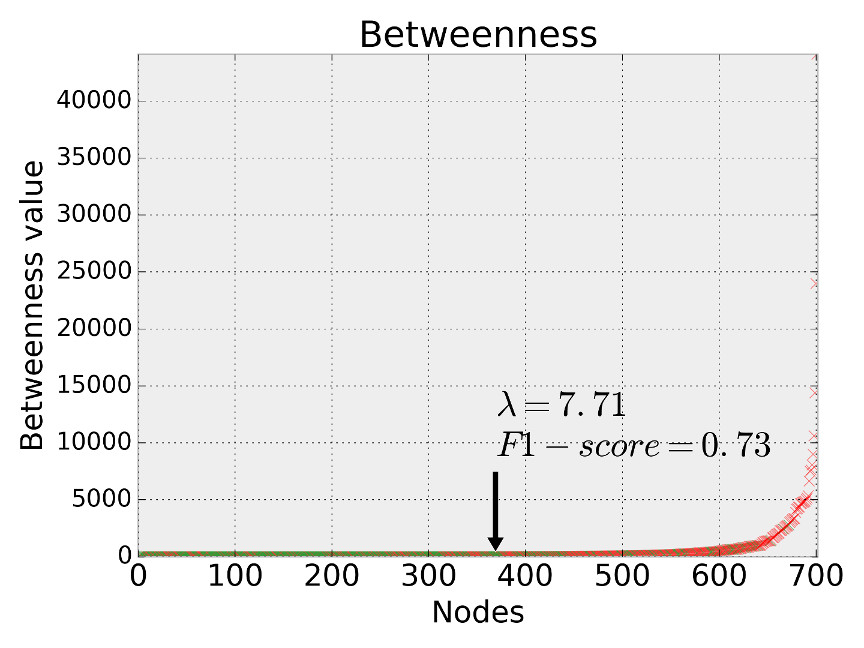}
\label{fig:threshold_betweenness}}
\qquad
\begin{minipage}{3.5cm}
\includegraphics[width=4.2cm]{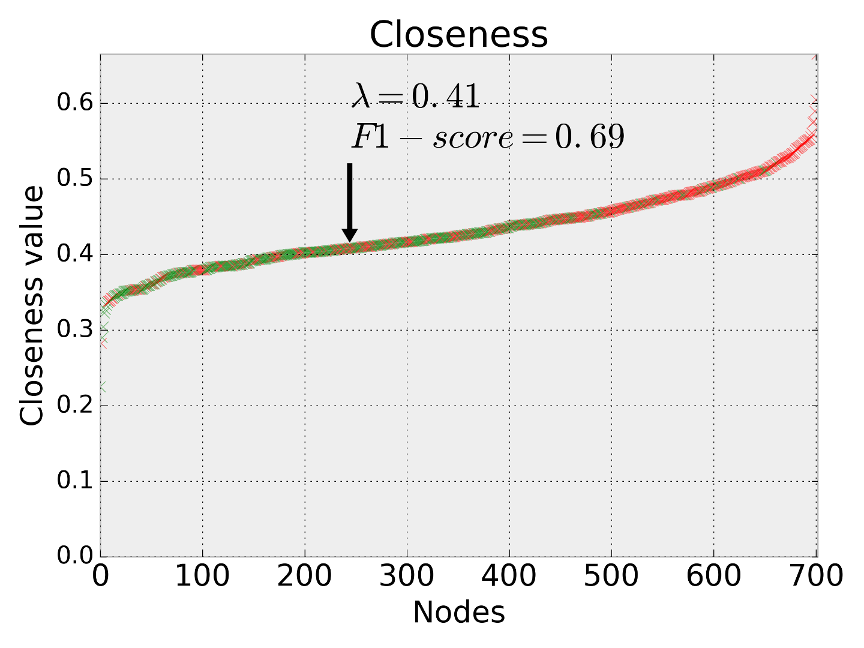}
\label{fig:threshold_closeness}
\end{minipage}
\qquad
\begin{minipage}{3.5cm}
\includegraphics[width=4.2cm]{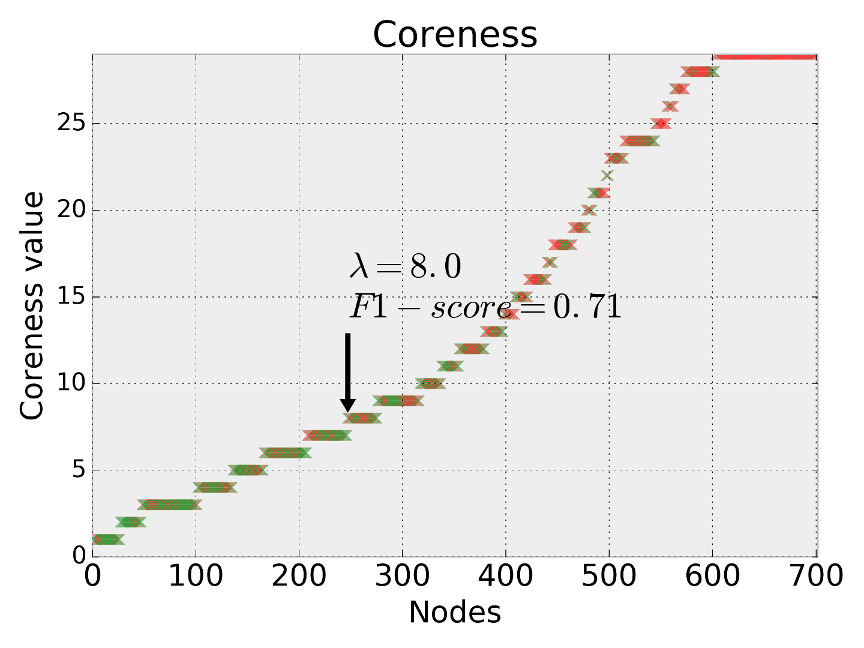}
\label{fig:threshold_coreness}
\end{minipage}
\qquad
\begin{minipage}{3.5cm}
\includegraphics[width=4.2cm]{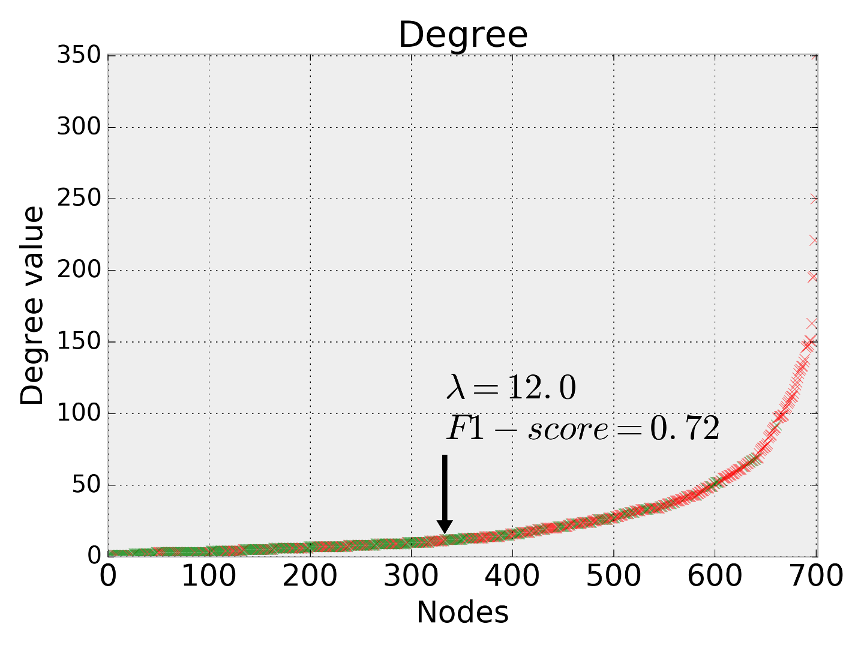}
\label{fig:threshold_degree}
\end{minipage}
\qquad
\begin{minipage}{3.5cm}
\includegraphics[width=4.2cm]{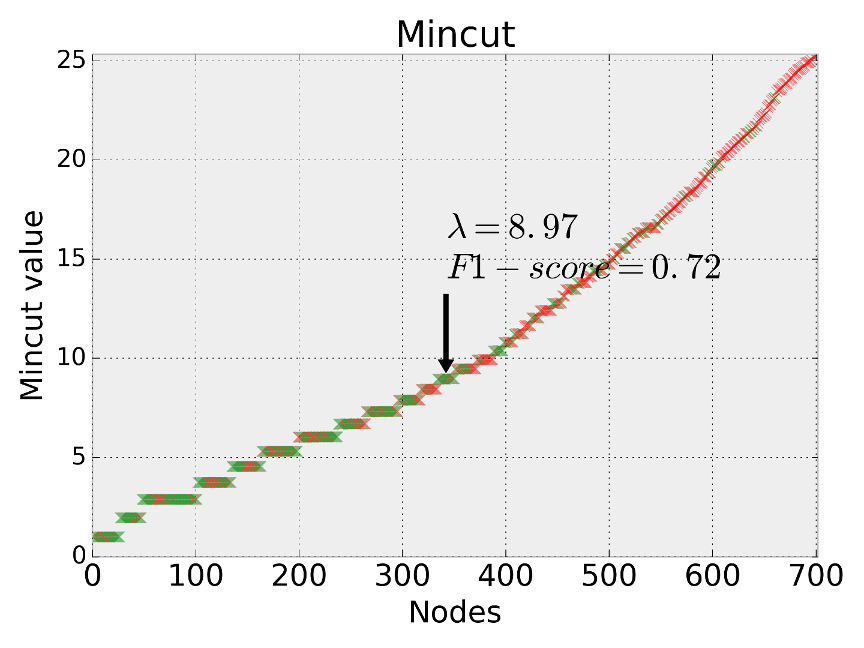}
\label{fig:threshold_minCut}
\end{minipage}
\qquad
\begin{minipage}{3.5cm}
\includegraphics[width=4.2cm]{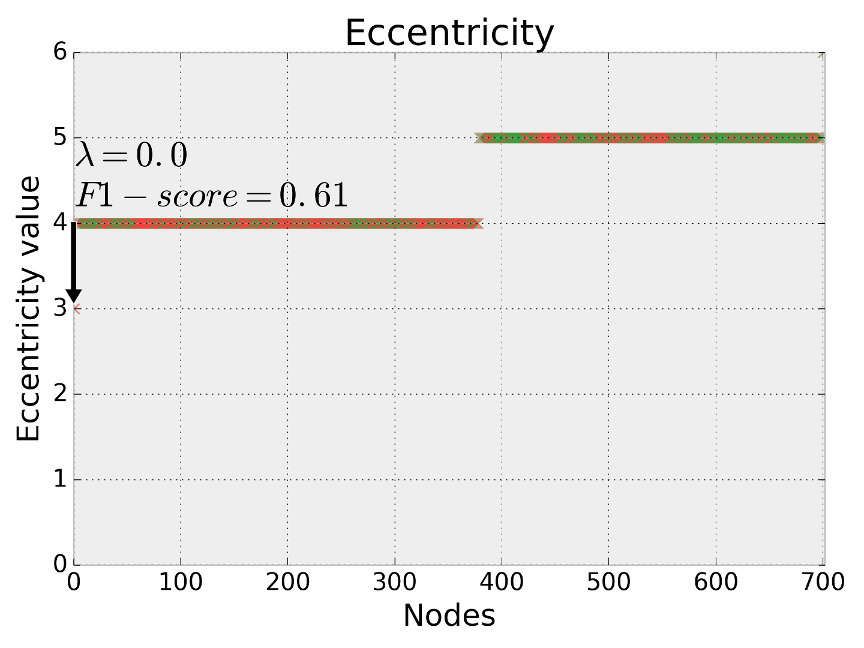}
\label{fig:threshold_eccentricity}
\end{minipage}
\qquad
\begin{minipage}{3.5cm}
\includegraphics[width=4.2cm]{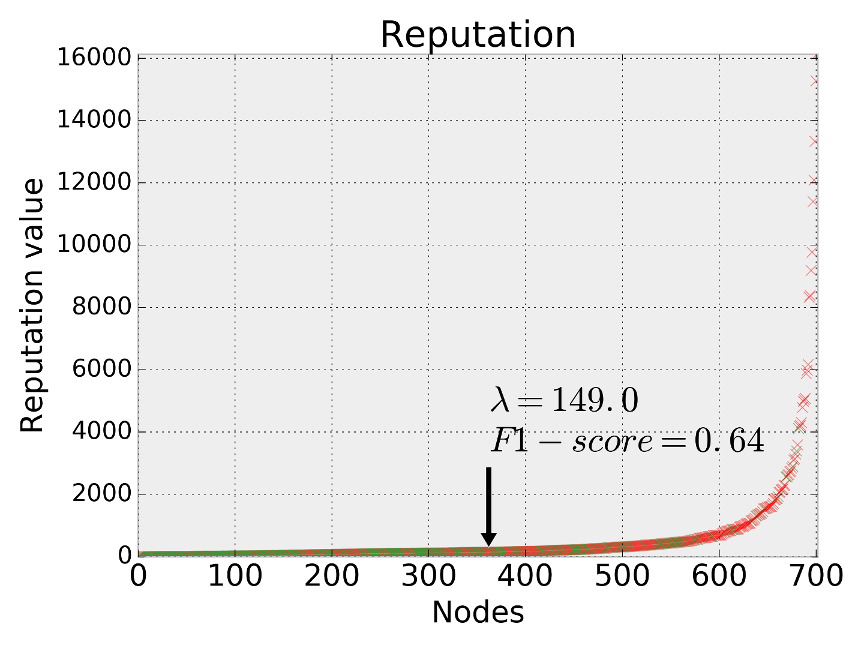}
\label{fig:threshold_reputation}
\end{minipage}
\qquad
\begin{minipage}{3.5cm}
\includegraphics[width=4.2cm]{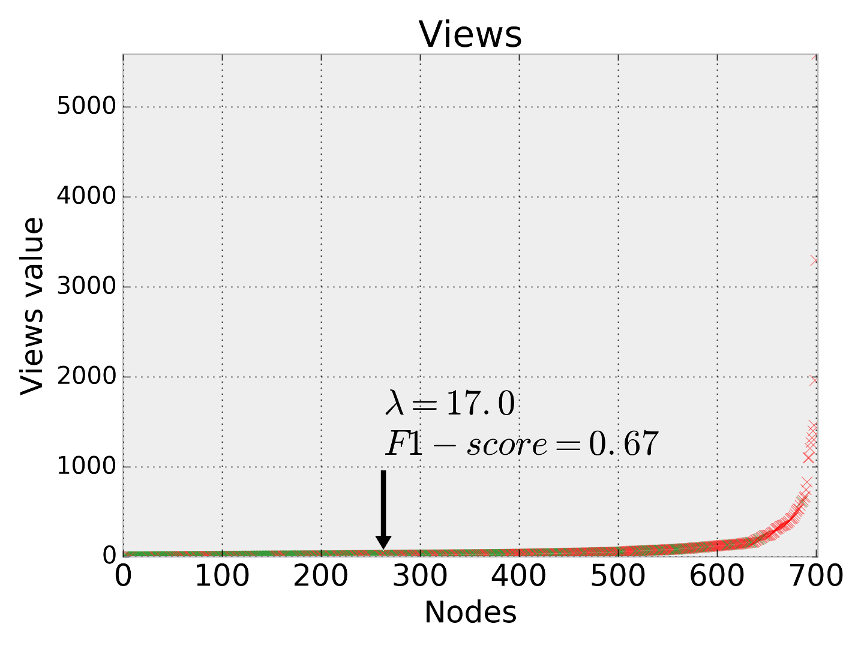}
\label{fig:threshold_views}
\end{minipage}
\qquad
\begin{minipage}{3.5cm}
\includegraphics[width=4.2cm]{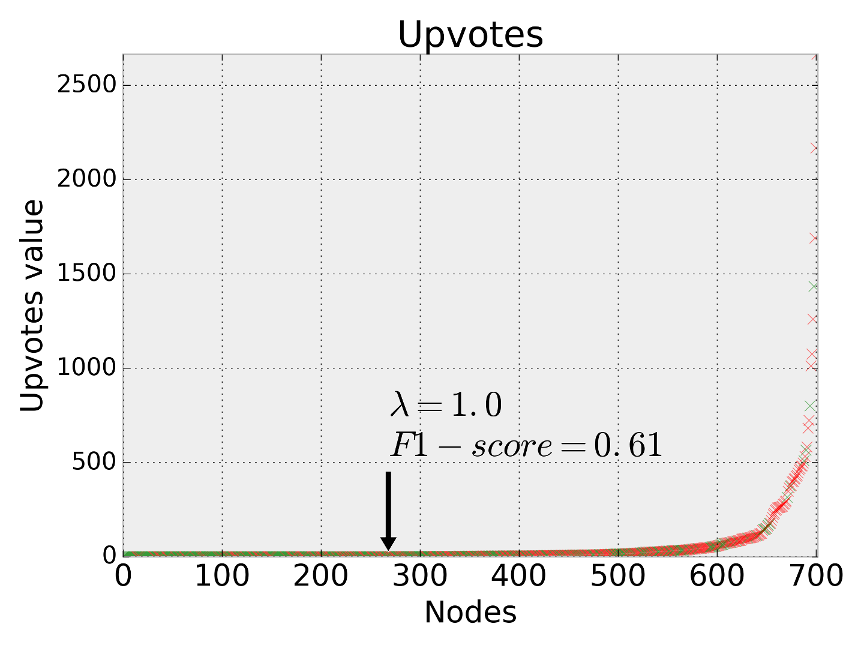}
\label{fig:threshold_upVotes}
\end{minipage}
\caption{(Color online) The figure shows the results of the STM in the training phase. We used the networks of the \textit{Business Startups} decayed dataset. The training period was from $Nov$-$2009$ to $Jan$-$2010$ to estimate the best $\lambda$ which was used on the test period from $Jan$-$2010$ to $Mar$-$2010$. The best value of the threshold $\lambda$, is shown in the figure, for each attribute associated with the value of the F1-score of the testing phase. The red and green markers indicate that the node did not leave or did, respectively. The x-axis represents the nodes ranked according to their attribute value.}
\label{fig:thresholdmodelprediction}
\end{figure*}

\begin{figure*}[!ht]
\centering
\parbox{3.5cm}{
\includegraphics[width=4.2cm]{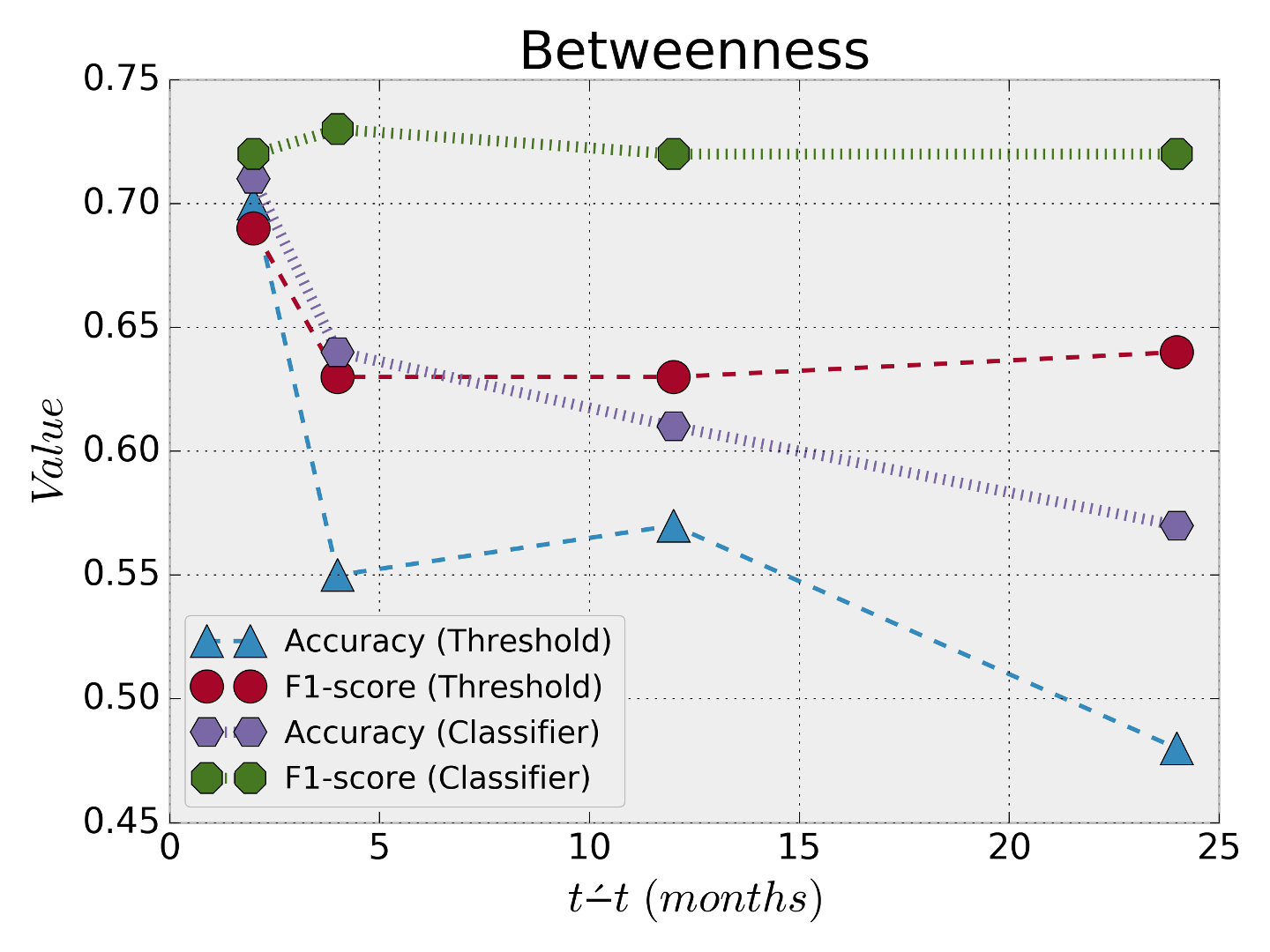}
\label{fig:betweenness}}
\qquad
\begin{minipage}{3.5cm}
\includegraphics[width=4.2cm]{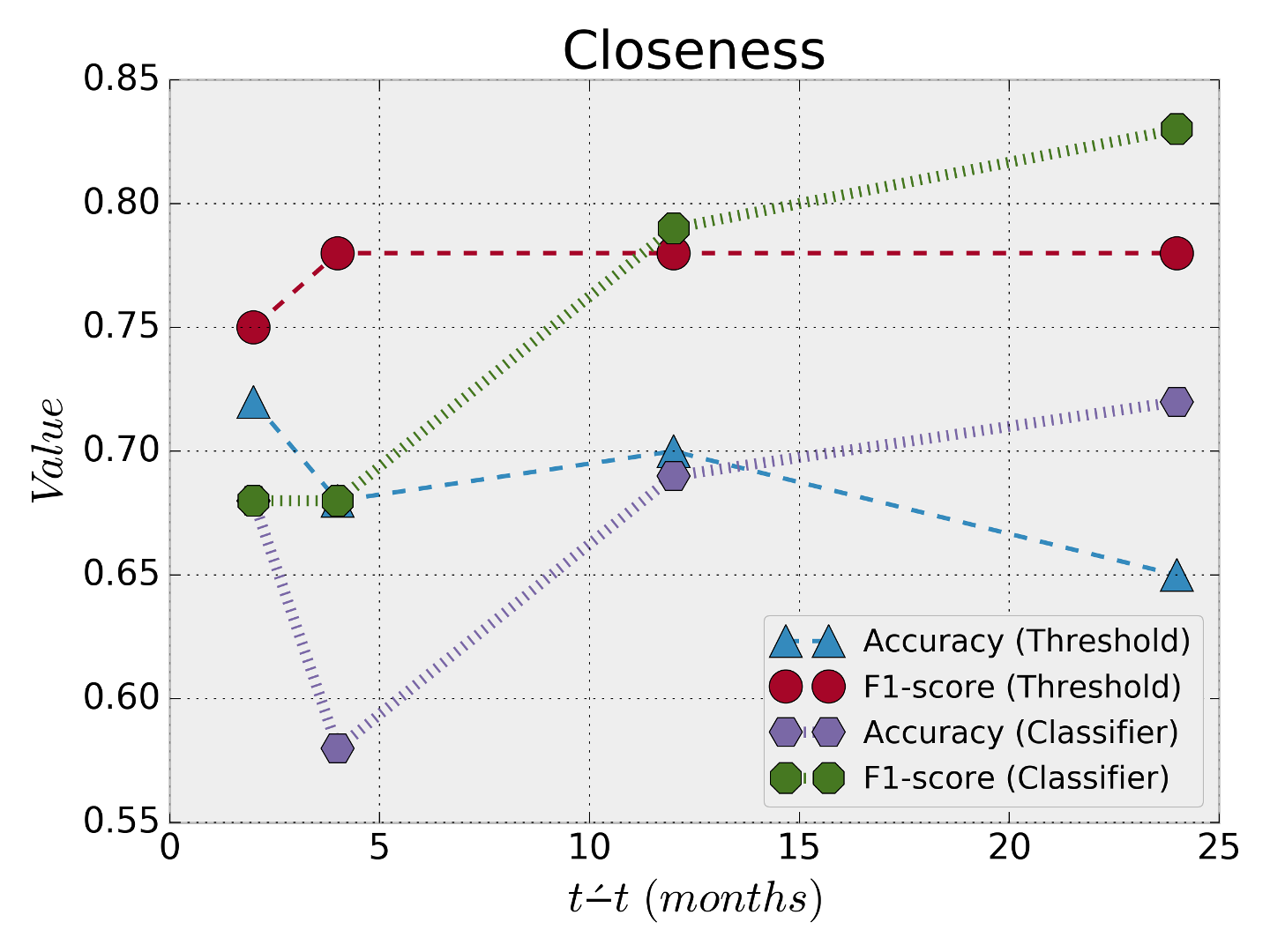}
\label{fig:closeness}
\end{minipage}
\qquad
\begin{minipage}{3.5cm}
\includegraphics[width=4.2cm]{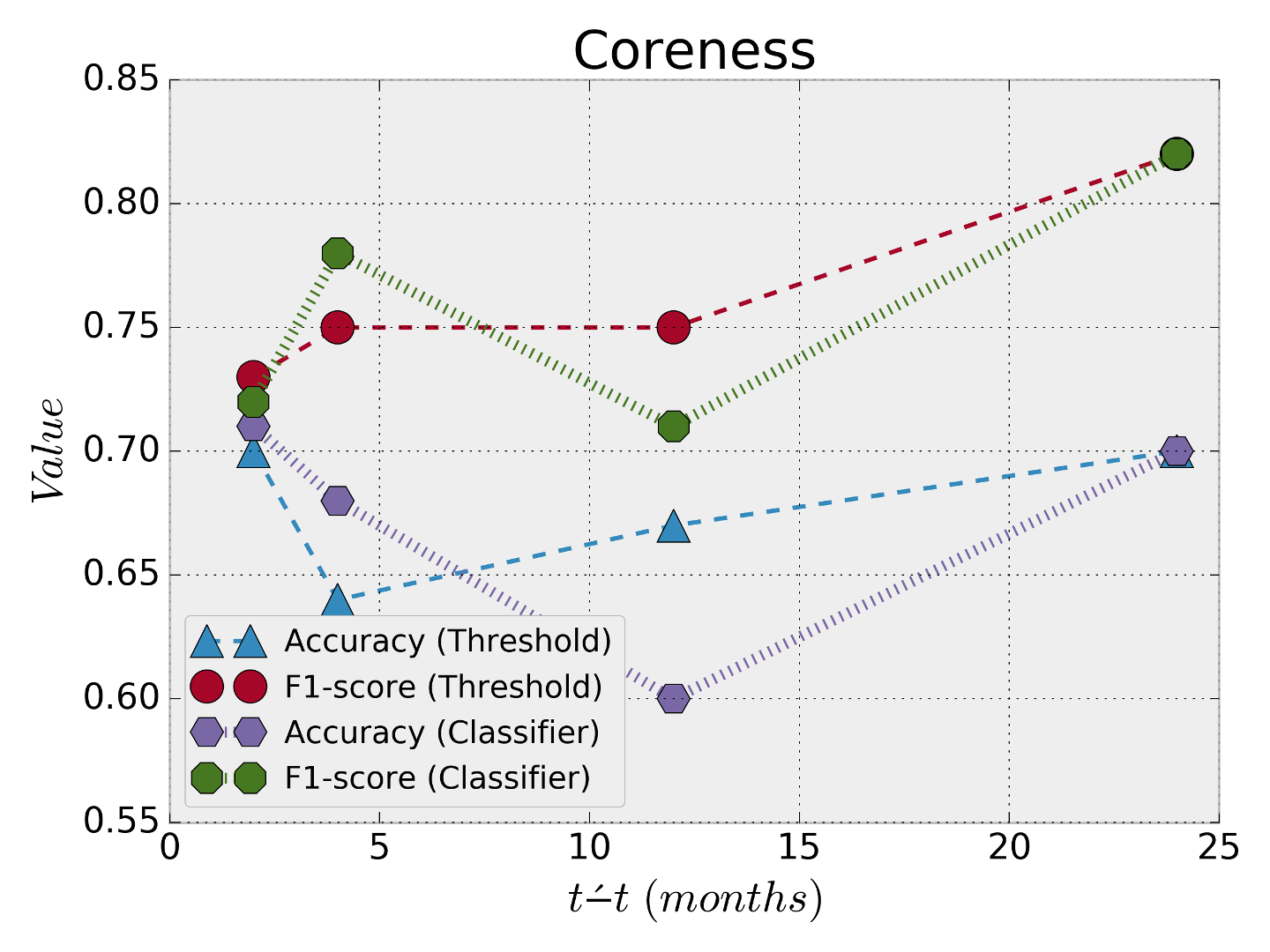}
\label{fig:coreness}
\end{minipage}
\qquad
\begin{minipage}{3.5cm}
\includegraphics[width=4.2cm]{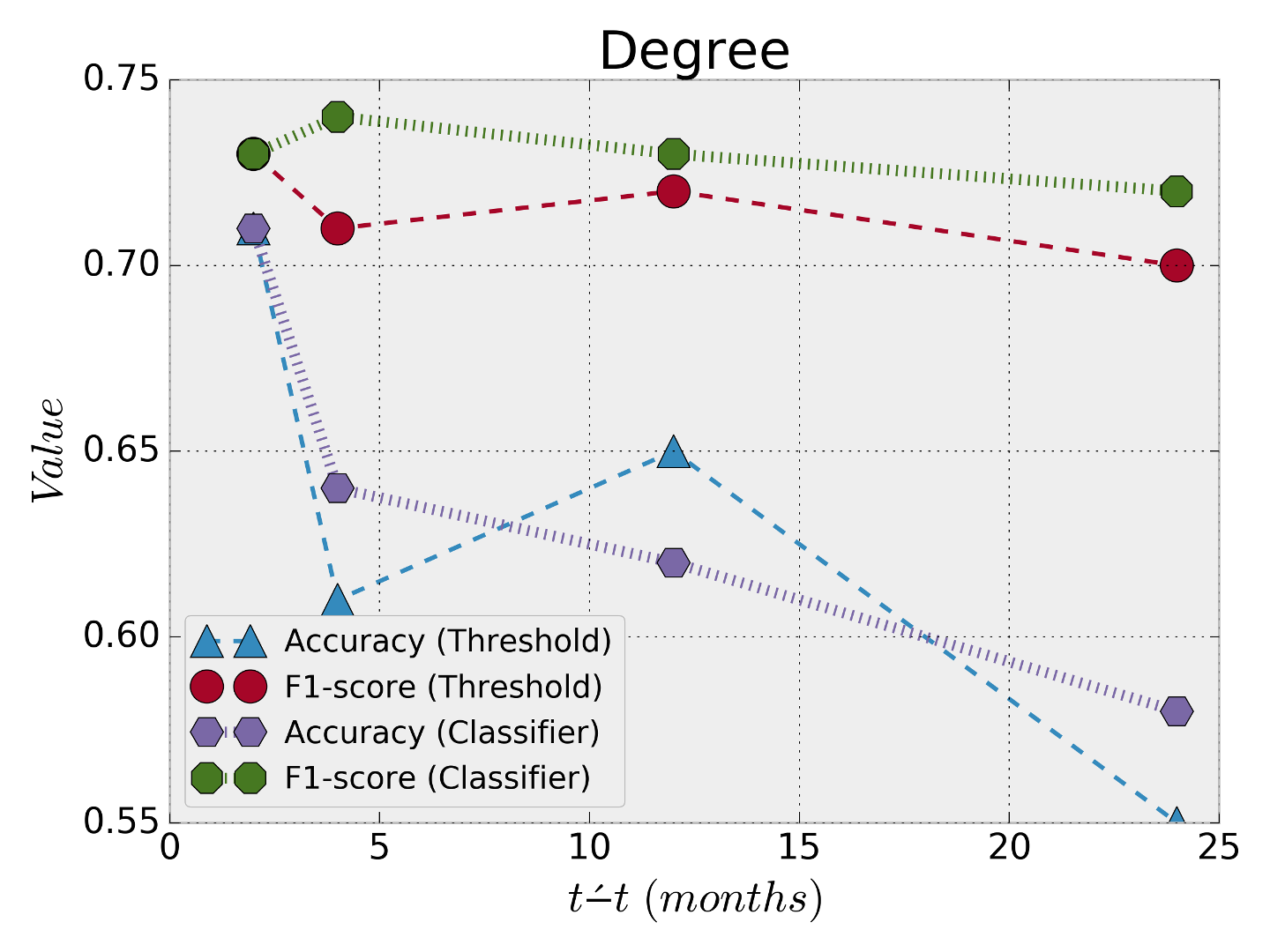}
\label{fig:degree}
\end{minipage}
\qquad
\begin{minipage}{3.5cm}
\includegraphics[width=4.2cm]{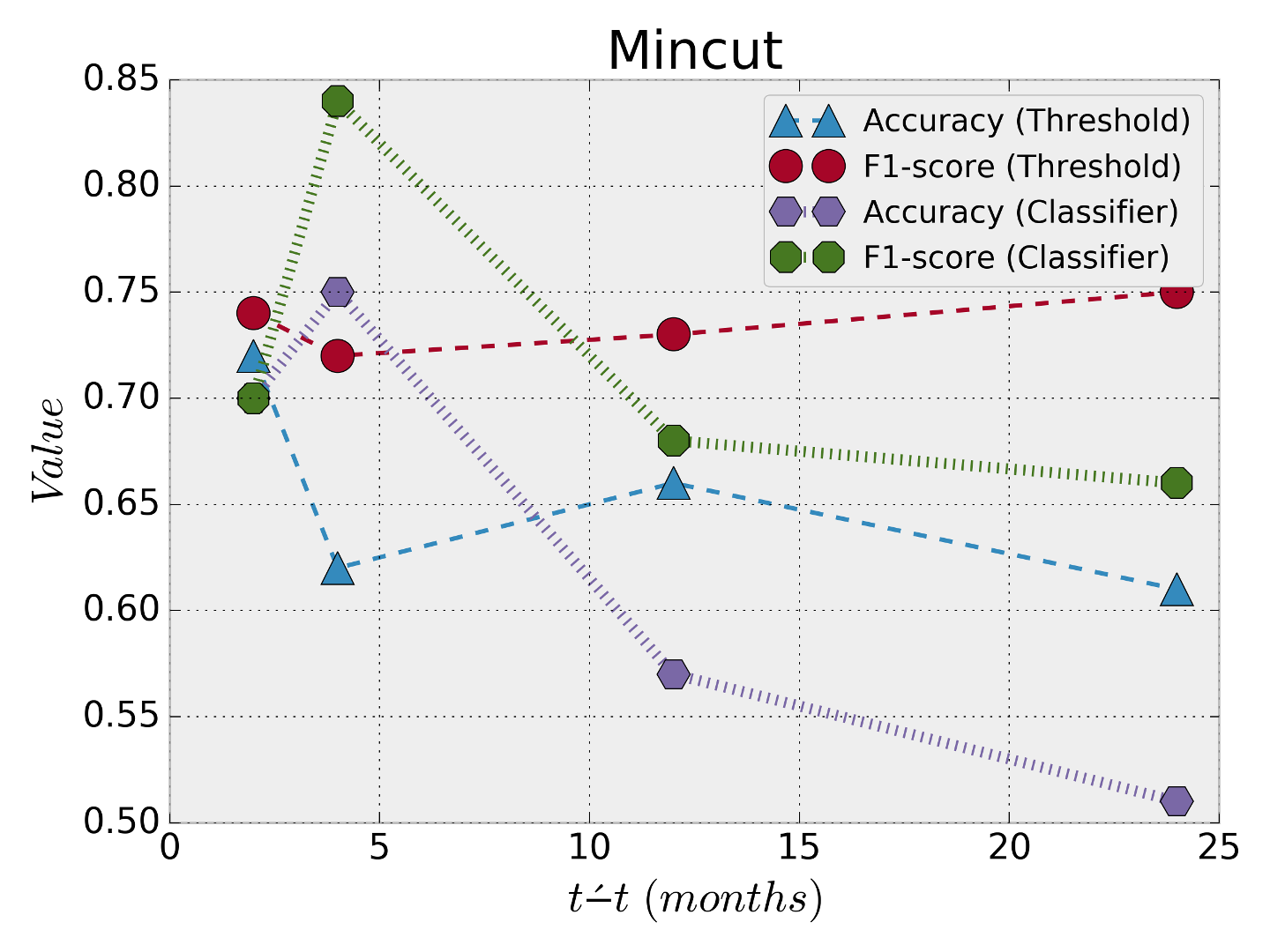}
\label{fig:minCut}
\end{minipage}
\qquad
\begin{minipage}{3.5cm}
\includegraphics[width=4.2cm]{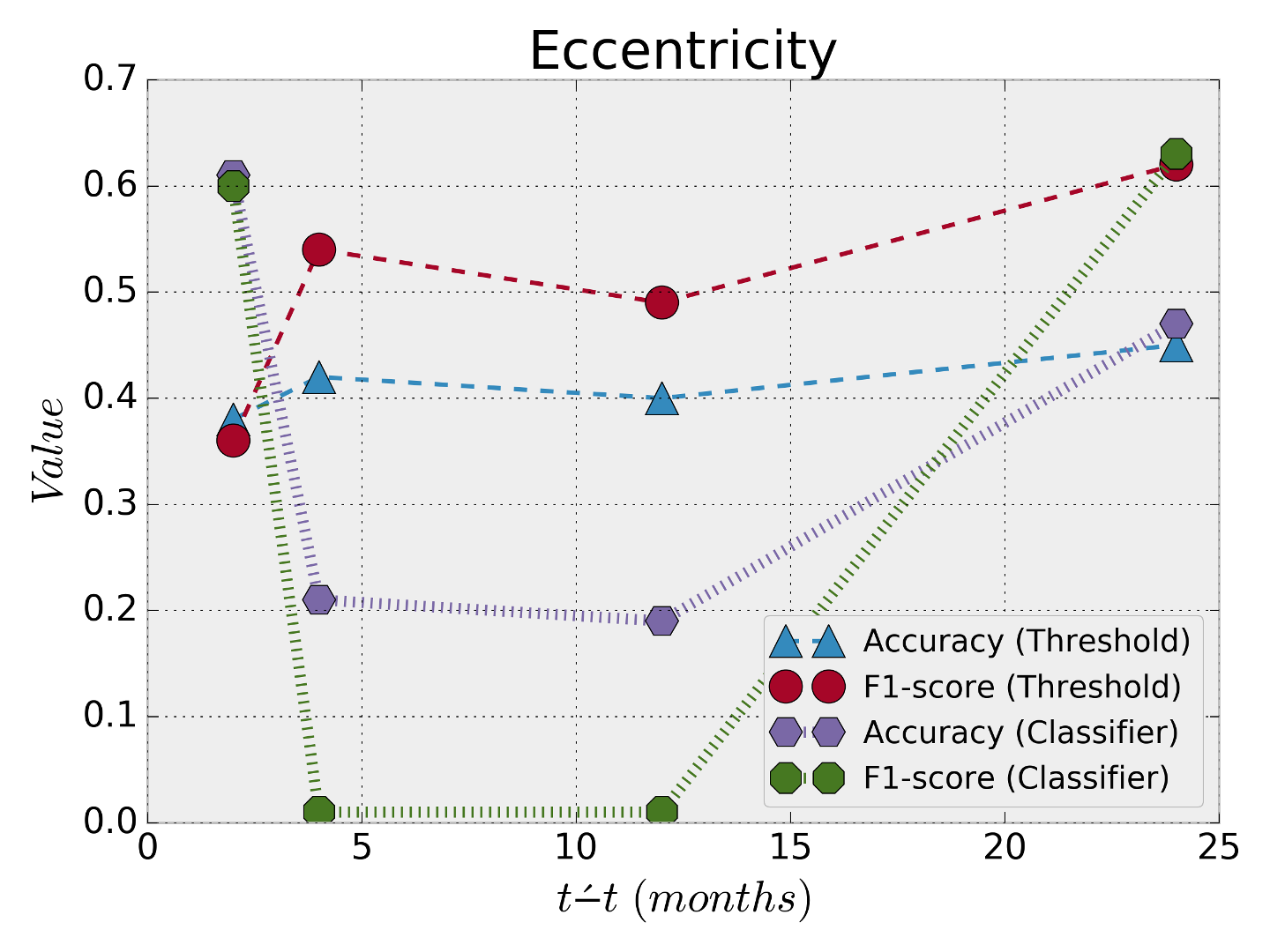}
\label{fig:eccentricity}
\end{minipage}
\qquad
\begin{minipage}{3.5cm}
\includegraphics[width=4.2cm]{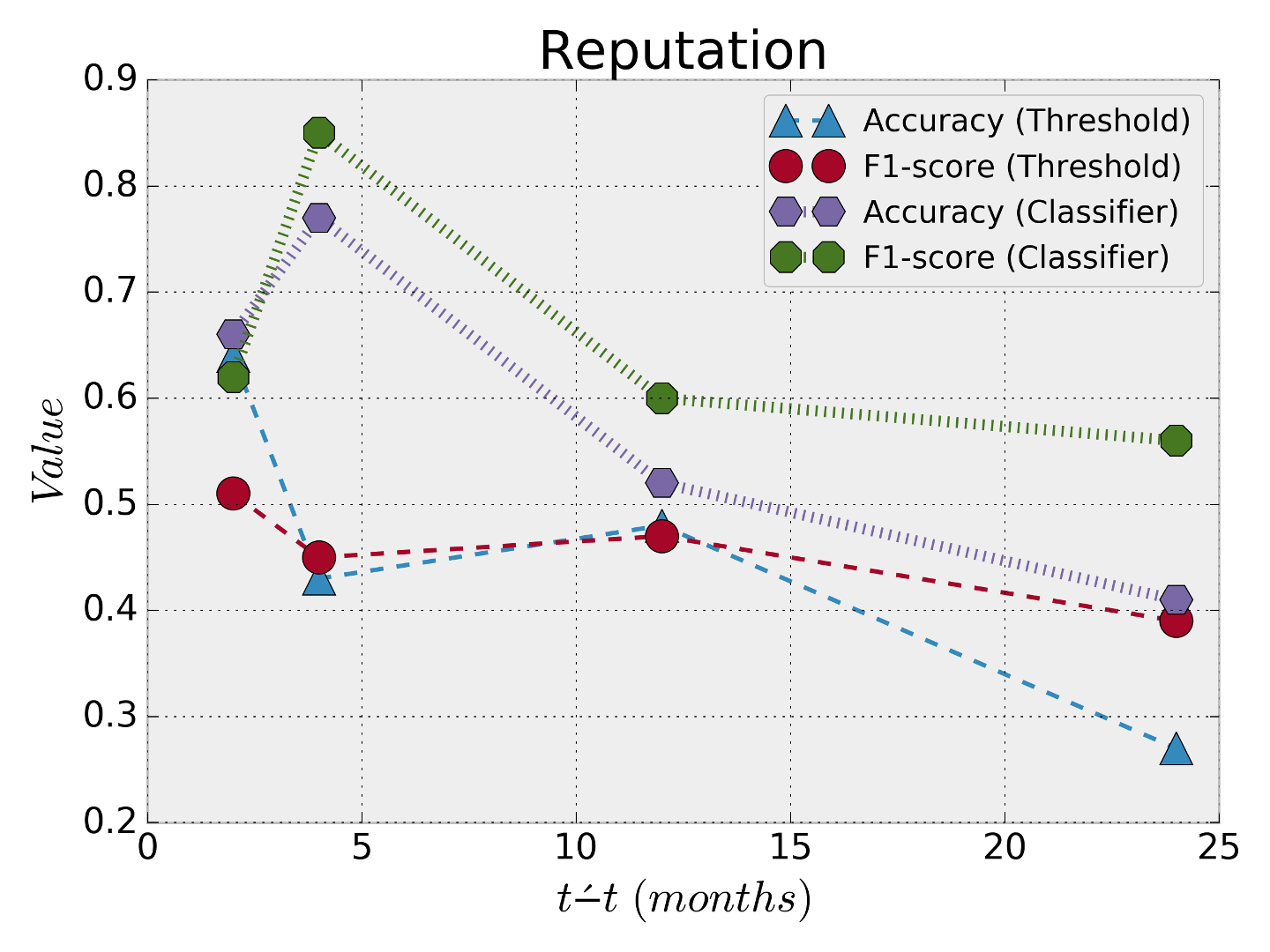}
\label{fig:reputation}
\end{minipage}
\qquad
\begin{minipage}{3.5cm}
\includegraphics[width=4.2cm]{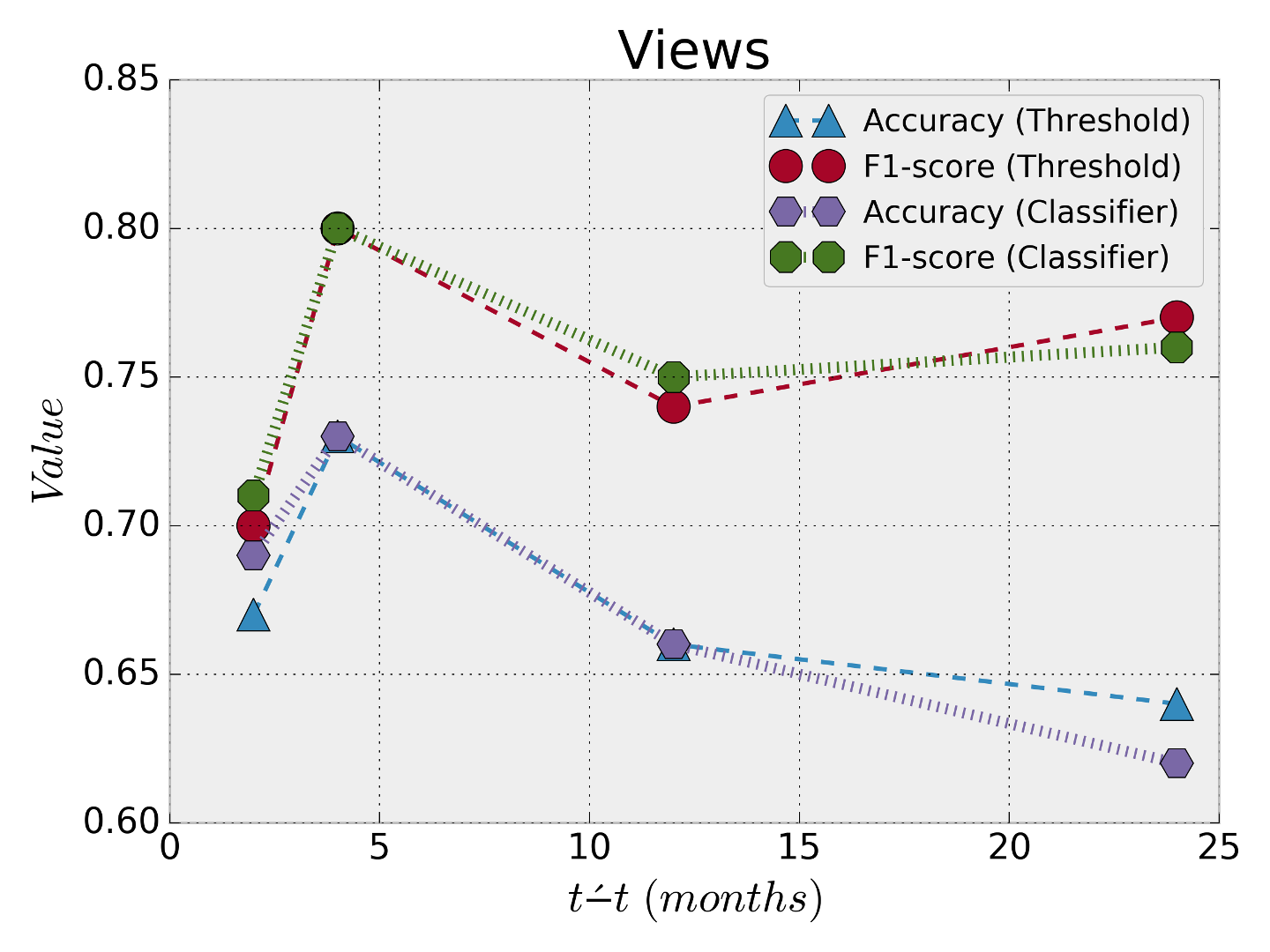}
\label{fig:views}
\end{minipage}
\qquad
\begin{minipage}{3.5cm}
\includegraphics[width=4.2cm]{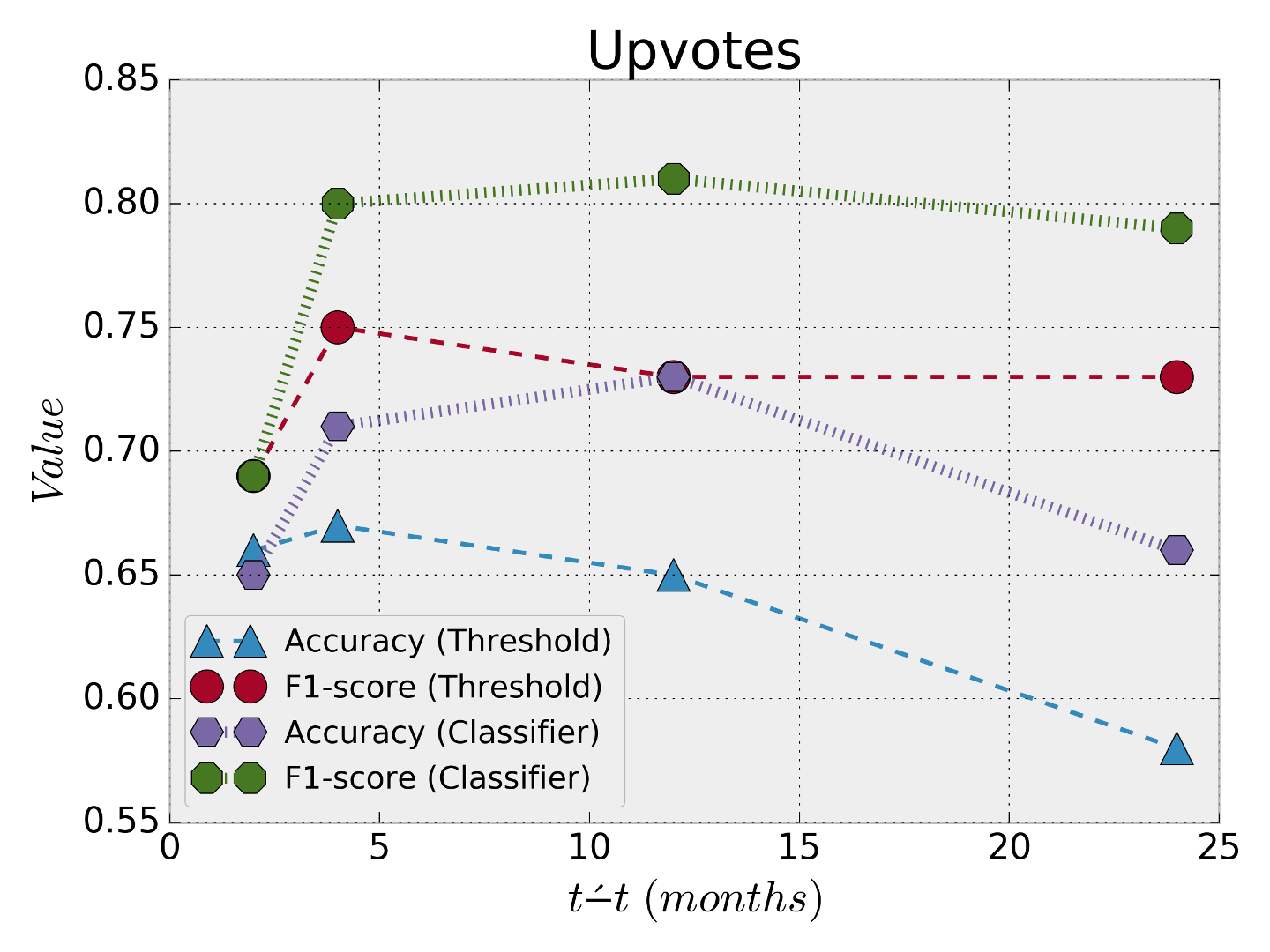}
\label{fig:upVotes}
\end{minipage}
\caption{(Color online) The figure shows the prediction performance in terms of the F1-score and the accuracy for the prediction using only one attribute. The figure shows the results for the prediction using the STM and the machine learning classification presented in Sections~\ref{subsec:simplethresholdmodel} and~\ref{subsec:machinelearningclassification}, respectively. We used the networks of the \textit{Business Startups} decayed dataset. The training period was done on the period $Nov$-$2009$ to $Jan$-$2010$ to estimate the best $\lambda$ which was used on the test periods $2,4,12,$ and $24$ months to get more insights regarding the prediction performance. Thus, the x-axis represents the prediction time in months and the y-axis represents the prediction measure values.}
\label{fig:performanceResults}
\end{figure*}

In this section, we provide the results of the community decay prediction using one attribute. We also provide the results of the machine learning binary classification using one attribute only. Thus, we can compare the performance of the STM with the machine learning model. Figure~\ref{fig:thresholdmodelprediction} shows the training results of the STM. The STM performs reasonably well for most of the attributes. For example, attributes like \textit{Betweenness}, \textit{Coreness}, \textit{Degree}, and \textit{Views} show an accepted F1-score. Some other attributes like the \textit{Upvotes} and the \textit{Eccentricity} contain no significant information to provide good prediction as their $\lambda$ was $zero$. 
Having trained the STM and obtained the corresponding $\lambda$ for each attribute, we then predict using the attribute value at $\lambda$ for different future time points. Figure~\ref{fig:performanceResults} shows the prediction results of the STM and also the machine learning model of one attribute. To our surprise, the performance of the STM was satisfactory.
For the attribute \textit{Closeness}, the STM outperforms the machine learning model slightly with an advantage of $0.02$ and $0.03$ of the accuracy and the F1-score, respectively, averaged over prediction periods $2,4,12,$ and $24$ months. A similar advantage was found on the attributes \textit{MinCut} and the \textit{Eccentricity}. On the other hand, there was a slight advantage for the machine learning model over the STM for the attributes \textit{Reputation}, \textit{Degree}, \textit{Betweenness}, and \textit{Upvotes}.
For example, using the attribute \textit{Betweenness}, the machine learning model outperforms the STM with only $0.05$ and $0.07$ in terms of the accuracy and the F1-score, respectively, averaged over the prediction periods $2,4,12,$ and $24$ months. Other attributes such as the \textit{Coreness} and the \textit{Views} show no difference on the averaged prediction sums over the different periods.
It is worth to mention that, the best prediction result of the STM was using the \textit{Coreness} attribute with prediction performance $0.83$ and $0.7$ for the F1-score and the accuracy, respectively, for the prediction of $24$ months. Also, the machine learning model's best accuracy and the F1-score was using the attribute \textit{Reputation} with value of $0.85$ and $0.73$, respectively. 
\subsection{Prediction using multiple attributes}
\label{subsec:predictionusingmultipleattributes}
In this section, we provide the prediction results of our machine learning framework. We used machine learning because we may lose a lot of information when limiting our prediction to only one attribute. We emphasis here that all of the experiments performed in this section were performed on two different datasets: one for the training phase, and the other for the testing phase, which supports validity of our results and conclusions which eliminates overfitting.
\subsubsection{Features properties:}
\label{subsubsec:featureproperties}
\begin{figure}
  \centering
    \includegraphics[scale=0.8]{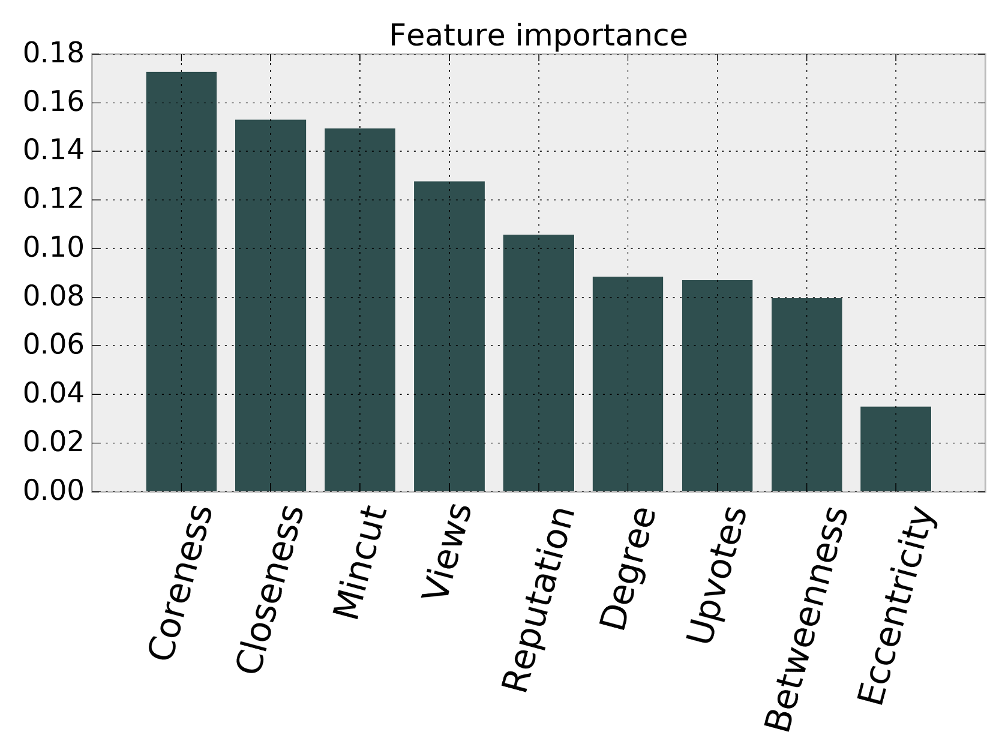}
  \caption{ The figure shows the feature importance such that $\sum_i^{i \in \vartheta \cup \theta}w(i) = 1$, where $w(i)$ is the feature importance. The used method for generating the importance is random forests where the importance of a feature increases whenever the split in the tree using that feature minimizes the prediction error~\cite{louppe2013}.}
  \label{fig:featureImportance}
\end{figure}

\begin{figure}[][h]
  \centering
    \includegraphics[scale=0.9]{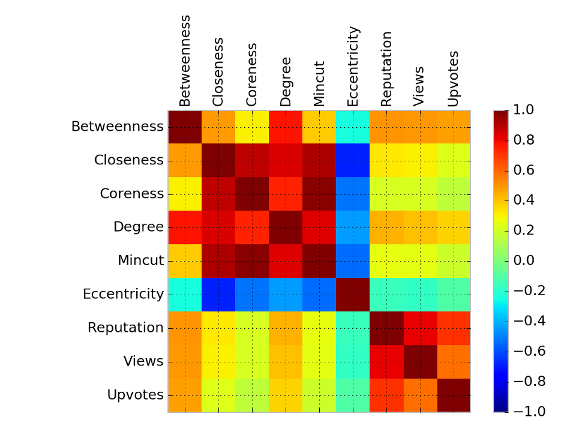}
  \caption{(Color online) The figure shows the Pearson's correlation coefficient values for the used features and it is defined as: $\rho(f_1,f_2) = \frac{Covariance(f_1,f_2)}{\sqrt{Variance(f_1) \cdot Variance(f_2)}}$, where $f_1, f_2 \in \vartheta \cup \theta$ and $\rho(f_1,f_2) \in [-1,1]$. The data used to generate this figure is the \textit{Business Startups} for the period between $Jan$-$2010$ and $Mar$-$2010$.}
  \label{fig:correlation_matrix}
\end{figure}

\begin{figure*}[!ht]
  \centering
    \includegraphics[scale=0.9]{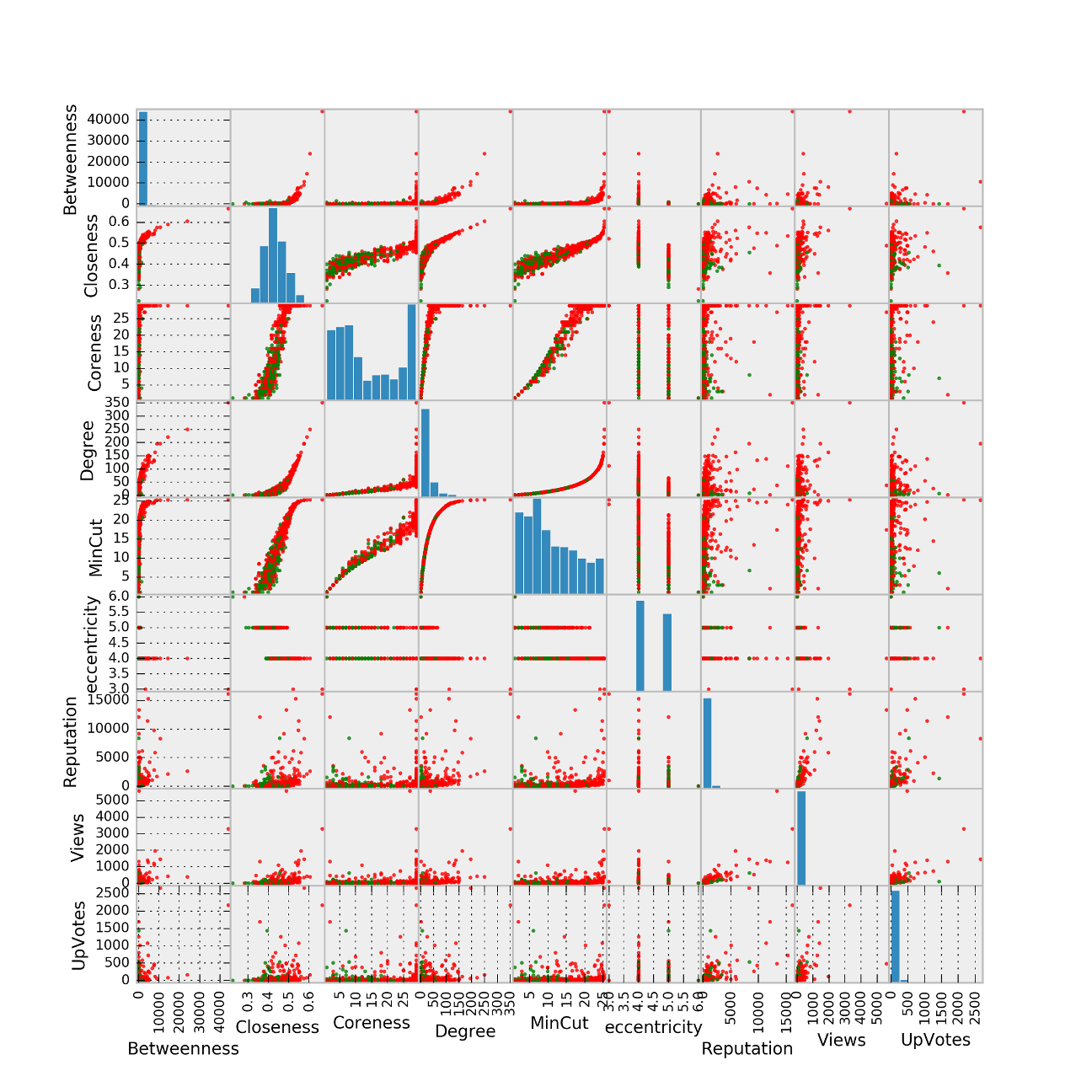}
  \caption{(Color online) The figure shows the distribution of each feature (in the diagonal plots) of the used model and also shows the correlation plot between each two attributes. The green points are nodes who left the network and the green ones are who did not. The data used to generate this figure is the \textit{Business Startups} for the period between $Jan$-$2010$ and $Mar$-$2010$.}
  \label{fig:correlation_scatter}
\end{figure*}

We started by investigating the properties of the used features. Ranking features according to their importance is vital in selecting the best attributes during training the testing phases. Figure~\ref{fig:featureImportance} shows the importance of the features used in this work. 
The figure shows that the \textit{Coreness} and the \textit{Closeness} are the most important network-based features and the \textit{Views} and the \textit{Reputation} are the most important exogenous information features. The information provided in this figure is valuable in selecting the best set of features. Thus, we provide different training and testing variations of the feature model $\mathcal{M}$ as follows:\\
\textbf{(1)} $\mathcal{M}$(all), which uses all features.\\
\textbf{(2)} $\mathcal{M}$(Best$_4$), which uses the best $4$ features based on Figure~\ref{fig:featureImportance}.\\
\textbf{(3)} $\mathcal{M}$(Best$_1$), which uses the best feature from the network-based features and the best feature from the exogenous attributes.\\
\textbf{(4)} $\mathcal{M}$(Best$_2$), which uses the best two features from the network-based features and the best two features from the exogenous attributes.\\
Using the set of all features is not always the best choice due to some inherited properties of the machine learning classifiers. For example, some classifiers are sensitive to correlated attributes and many classifiers perform poorly with low variance attributes. Thus, we provide additional analysis of the attributes in order to better understand the features. Figure~\ref{fig:correlation_matrix} shows the Pearson's correlation coefficient matrix of the attributes. Values close to $-1$ indicate a negative correlation, while values near to $1$ indicate a positive correlation. It is preferable to feed machine learning classifiers with as more uncorrelated features as possible. We notice from Figure~\ref{fig:correlation_matrix} that the exogenous features are more correlated to each others. Also, the network-based attributes are more correlated to each other. To see that, Figure~\ref{fig:correlation_scatter} provides more information about distribution of the attributes along with one-to-one scatter plot.  For example, we notice that \textit{Coreness} vs \textit{Degree} and \textit{Closeness} vs \textit{MinCut} are highly correlated. 

\begin{figure*}[]
  \centering
    \includegraphics[scale=0.45]{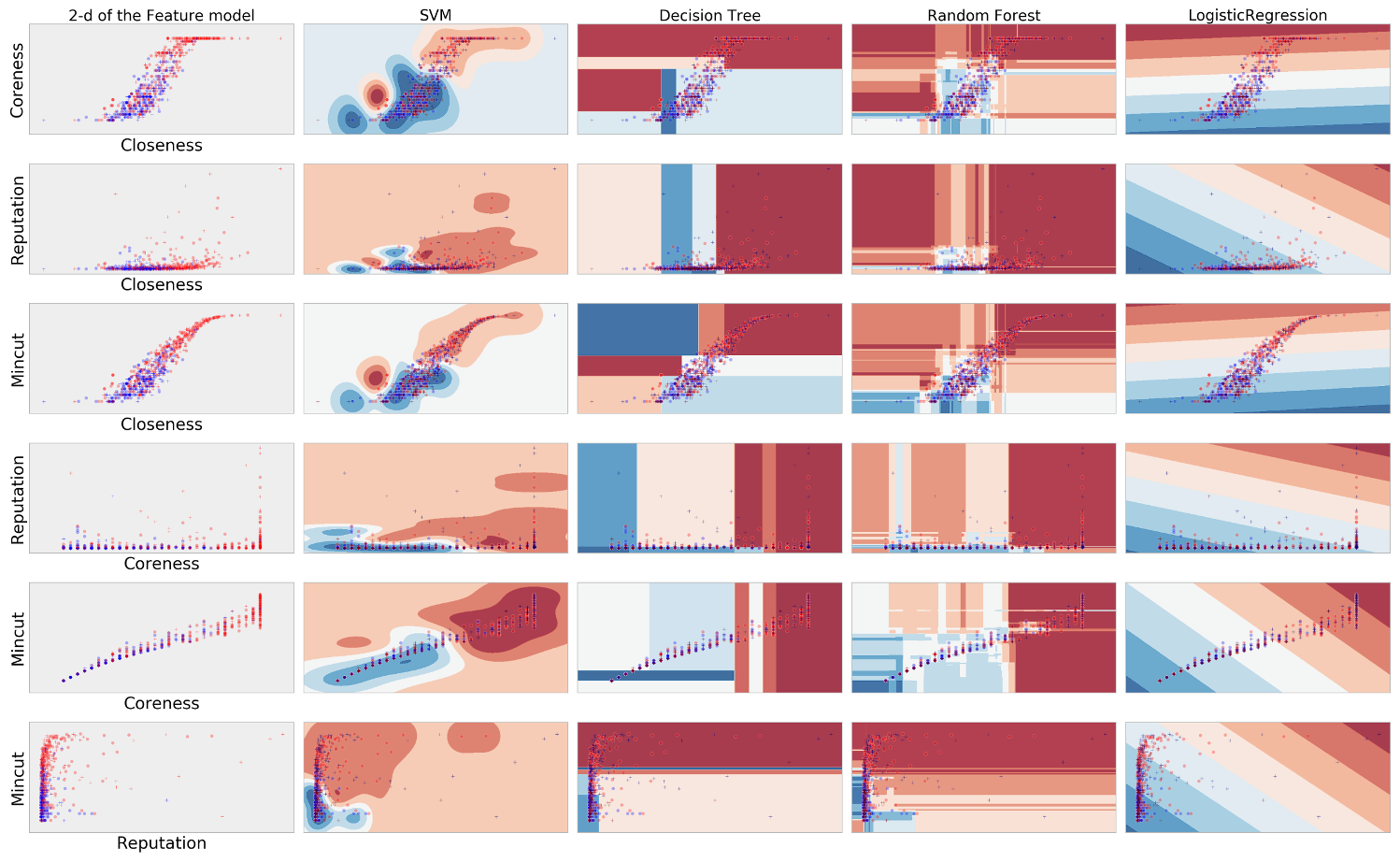}
  \caption{(Color online) The figure shows how able are the machine learning classifiers to segregate non-linearly separable data. We show a $2$-$d$ of the attributes of the \textit{Business Startups}. The figures in most left column show two features' values as a scatter plot. The blue points represent the nodes who left and the red points represent the nodes who did not leave. The points marked with solid circle are the points of the training set and the points marked with '+' are points of testing set. Each of the other plots in the other columns represent how different probabilistic classifier managed to classify the corresponding $2$-$d$ dataset. The color gradient in the prediction plots is the probability of the prediction, i.e., the darker the color, the closer the probability to $1$ or $0$ (where $1$ means left and $0$ means did not leave).} 
  \label{fig:decisionBoundaries}
\end{figure*}

Figure~\ref{fig:correlation_scatter} shows also the distribution of each attribute in the diagonal. The features \textit{Closeness}, \textit{Coreness}, \textit{Degree}, and \textit{MinCut} have more variance than the others, which gives a interpretation on why these attributes got more importance in Figure~\ref{fig:featureImportance}.
The data shown in Figure~\ref{fig:thresholdmodelprediction} are clearly non-linearly separable, i.e, there exist no possible threshold that separates the green and the red points. The separation of the data becomes even harder when incorporating more features, like the data points in Figure~\ref{fig:correlation_scatter}. In Figure~\ref{fig:decisionBoundaries}, we show an exemplary prediction on a $2$-$d$ data of the used attributes. The Figure illustrates how classifiers, such as the support vector machines, are able to provide a smooth probabilistic areas for separating points. For example, the blue and the red points of the \textit{MinCut} vs \textit{Closeness} are very interweaving, and the SVM managed to find good separation areas compared to the Logistic regression, Decision Trees, and Random forest. That's why we resort basically to the SVMs in the following results\footnote{The technical details of the SVMs can be found here~\cite{svm1995}.}.
\subsubsection{Prediction results:}
\label{subsubsec:predictionresults}
In this section, we provide the prediction results of the machine learning prediction framework. Table \ref{tab:resultsBusinessstartups} shows the prediction results of the \textit{Business Startups} decayed dataset using the \textit{four} variations of the framework and for \textit{four} time periods. The prediction results overall are satisfactory and the variation \textit{ Best}$_2$ gave the best prediction results over the other variations for all prediction periods. One thing to notice is that, when the prediction time grows in the prediction period, the prediction performance increases. One interpretation is that the machine learning classifiers were able to learn the leave patterns much more than the stay pattern, given that the decayed communities networks ends with a disconnected networks, i.e., all of its nodes had already left the network. Similar results were found for the \textit{Literature} dataset in Table~\ref{tab:resultslit}. The prediction results of the \textit{Literature} dataset are higher than the dataset of \textit{Business Startups}. After investigating the two datasets, we found the \textit{Business Startups} community went through two phases of decay. After the first decay of the community website, there was another relaunch which was also not successful and ended with second decay. This interprets its long time span compared to other decayed communities. Thus, there was a fluctuation in the activity of that community which made it harder for the classifiers to identify the real leave patterns in this community. Table~\ref{tab:resultslit} shows no significant difference between the \textit{four} variations of the model, except the variation Best$_1$ which shows a slight advantage over the other variations in the prediction period $2$ months.

\begin{table}[!ht]
\centering
\footnotesize
\begin{tabular}{|c|c|c|c|c|c|c|c|c|}
\hline
\multirow{2}{*}{Attributes}                     & \multicolumn{2}{c|}{2 Months} & \multicolumn{2}{c|}{4 Months} & \multicolumn{2}{c|}{12 Months} & \multicolumn{2}{c|}{24 Months} \\ \cline{2-9} 
                                                & $\mathcal{A}$           & F1            & $\mathcal{A}$           & F1            & $\mathcal{A}$           & F1             & $\mathcal{A}$           & F1             \\ \hline
$\mathcal{M}$(All)                                          & 0.72               & 0.74          & 0.68               & 0.79          & 0.69               & 0.79           & 0.76               & 0.85           \\ \hline
$\mathcal{M}$(Best$_4$)         & 0.73               & 0.74          & 0.81               & 0.72          & 0.68               & 0.77           & 0.72               & 0.82           \\ \hline
$\mathcal{M}$(Best$_1$)                                   & 0.74               & 0.73          & 0.66               & 0.73          & 0.66               & 0.76           & 0.7                & 0.81           \\ \hline
$\mathcal{M}$(Best$_2$)
 & 0.72               & 0.75          & 0.80                & 0.87          & 0.81               & 0.88           & 0.78               & 0.87           \\ \hline
\end{tabular}
\caption{The table shows the prediction results of the machine learning classifier for the networks constructed from the decayed \textit{Business Startups} community dataset. The training period was from $Nov$-$2009$ to $Jan$-$2010$. The table shows the prediction for different testing sets, namely after $2,4,12,$ and $24$ months. The prediction was done using different variations of the attributes model $\mathcal{M}$ as presented in Section~\ref{subsec:featuremodel}.}
\label{tab:resultsBusinessstartups}
\end{table}
\begin{table}[!ht]
\centering
\footnotesize
\begin{tabular}{|c|c|c|c|c|c|c|c|c|}
\hline
\multirow{2}{*}{Attributes}                     & \multicolumn{2}{c|}{2 Months} & \multicolumn{2}{c|}{4 Months} & \multicolumn{2}{c|}{8 Months} & \multicolumn{2}{c|}{12 Months} \\ \cline{2-9} 
                                                & $\mathcal{A}$           & F1            & $\mathcal{A}$           & F1            & $\mathcal{A}$           & F1            & $\mathcal{A}$           & F1            \\ \hline
$\mathcal{M}$(All)                                          & 0.68               & 0.77          & 0.83               & 0.91          & 0.82               & 0.9           & 0.88               & 0.94          \\ \hline
$\mathcal{M}$(Best$_4$)       & 0.7                & 0.76          & 0.83               & 0.91          & 0.82               & 0.9           & 0.88               & 0.94          \\ \hline
$\mathcal{M}$(Best$_1$)                                 & 0.82               & 0.85          & 0.83               & 0.91          & 0.82               & 0.9           & 0.88               & 0.94          \\ \hline
$\mathcal{M}$(Best$_2$) & 0.68               & 0.72          & 0.83               & 0.91          & 0.82               & 0.9           & 0.88               & 0.94          \\ \hline
\end{tabular}
\caption{The table shows the prediction results of the machine learning classifier for the networks constructed from the decayed \textit{Literature} community dataset. The training period was performed from $Aug$-$2011$ to $Sep$-$2011$. The table shows the prediction for different testing sets, namely after $2,4,12,$ and $24$ months for different variations of the attributes model $\mathcal{M}$ presented in Section~\ref{subsec:featuremodel}. }
\label{tab:resultslit}
\end{table}
\begin{table}[!ht]
	\centering
	\footnotesize
	\begin{tabular}{|c|c|c|c|c|c|c|c|c|c|c|}
		\hline
		\multirow{2}{*}{Attributes}                     & \multicolumn{2}{c|}{2 Months} & \multicolumn{2}{c|}{4 Months} & \multicolumn{2}{c|}{12 Months} & \multicolumn{2}{c|}{24 Months} & \multicolumn{2}{l|}{36 Months} \\ \cline{2-11} 
		& $\mathcal{A}$           & F1            & $\mathcal{A}$           & F1            & $\mathcal{A}$           & F1             & $\mathcal{A}$           & F1             & $\mathcal{A}$           & F1             \\ \hline
		$\mathcal{M}$(All)                                          & 0.71               & 0.7           & 0.7                & 0.71          & 0.73               & 0.79           & 0.65               & 0.74           & 0.6                & 0.73           \\ \hline
		$\mathcal{M}$(Best$_4$)      & 0.73               & 0.73          & 0.74               & 0.78          & 0.75               & 0.82           & 0.83               & 0.9            & 0.83               & 0.9            \\ \hline
		$\mathcal{M}$(Best$_1$)                                   & 0.73               & 0.71          & 0.74               & 0.77          & 0.75               & 0.82           & 0.82               & 0.89           & 0.81               & 0.89           \\ \hline
		$\mathcal{M}$(Best$_2$) & 0.71               & 0.67          & 0.72               & 0.80           & 0.74               & 0.82           & 0.81               & 0.88           & 0.88               & 0.94           \\ \hline
	\end{tabular}
	\caption{The table shows the prediction results of the machine learning classifier for the networks constructed from the \textbf{alive} "Latex" community dataset. The train period was performed for the period $Jun$-$2010$ to $Sep$-$2010$. The table shows the prediction for different testing sets, namely after $2,4,12,24,$ and $36$ months for different variations of the attributes model $\mathcal{M}$ presented in Section~\ref{subsec:featuremodel}. Being alive, the Latex community made it possible to predict using $36$ months. }
	\label{tab:resultslatex}
\end{table}
\begin{table}[!ht]
	\centering
	\footnotesize
	\begin{tabular}{|c|c|c|c|c|c|}
		\hline
		\multicolumn{2}{|c|}{Datasets}        & \multicolumn{2}{c|}{4 Months} & \multicolumn{2}{c|}{8 Months} \\ \hline
		Train on (Decayed)                & Test on (Alive)    & $\mathcal{A}$          & F1            & $\mathcal{A}$          & F1            \\ \hline
		\multirow{2}{*}{Business Startups} & Latex      & 0.83               & 0.89          & 0.88               & 0.94          \\ \cline{2-6} 
		& Statistics & 0.79               & 0.84          & 0.79               & 0.81          \\ \hline
		\multirow{2}{*}{Literature}     & Latex      & 0.72               & 0.74          & 0.80                & 0.89          \\ \cline{2-6} 
		& Statistics & 0.77               & 0.80           & 0.72               & 0.78          \\ \hline
	\end{tabular}
	\caption{Results of cross-datasets prediction where the training was performed on decayed communities and the test was performed on alive communities. We trained the machine learning classifier on the $\mathcal{M}$(Best$_2$), which provided the best results. }
	\label{tab:resultsCrossdatasets}
\end{table}
It was tempting to test the alive communities also. Hence, we used the \textit{Latex} alive community to predict the leave of their members also. The prediction performance for it was also satisfactory. Again, the variation Best$_1$ shows a slight advantage over the other variations, except for the period prediction over $36$ months, which is considered a long time window to predict as shown in Table~\ref{tab:resultslatex}.
Then, we predicted the future of the activity of the members of the active communities such as the \textit{Latex} and the \textit{Statistics} using the leave information of the decayed communities such as \textit{Business Startups} and \textit{Literature}. Table~\ref{tab:resultsCrossdatasets} shows the prediction results when training the classifiers on the datasets of decayed communities and test the classifier on the datasets of alive communities. The results suggest a better prediction performance when compared with the prediction on the same communities, such as the results in Table~\ref{tab:resultslatex}. For instance, the F1-score at time period $4$ months was $0.89$ when trained on a decayed dataset compared to $0.80$ when trained on the \textit{Latex} dataset itself. The prediction also shows satisfactory results for predicting the leave of the \textit{Statistics} community when learning from the decayed communities.\\
\vspace*{-5mm}
\section{Discussion}
\label{sec:discussion}
\vspace*{-2mm}
\subsection{Answering the research questions}
\label{subsec:answeringRQ}
\textbf{Discussion on \textit{RQ1}}, \textit{How efficient is it to predict members leaving a social community using network-based measures?}: Based on the previous presentation of the models and the results presented in Section~\ref{sec:results}, it is clear that using network-based attributes provides good prediction performance in terms of F1-score and the accuracy. The simple prediction model showed acceptable prediction results when using only one network-based measure. The results were even better and more robust when using multiple network-based attributes for the machine learning model. However, not all of the attributes were of equal quality for decay prediction. For example, the \textit{Eccentricity} measure was rather useless as it showed bad prediction performance using the STM. Even worse, that measure is misleading as it showed very high prediction for the $24$ months using the STM, which was only the case because its initial $\lambda$ was calculated as $zero$. It may be impossible to have both the network structure and exogenous information available, as one or other might be hard to collect. Thus, an interesting aspect regarding this research question is to test the prediction using only the attributes of either the network-based features or the exogenous features. Our preliminary results for this sub-question showed very close prediction performance for perdition using only one of the two types of the features. For the \textit{Business Startups}, the accuracy and the F1-score were $0.75$ and $0.83$, respectively, for the network-based measures compared to $0.77$ and $0.87$ for the exogenous features. These results are comparable and suggest that both types of features can be used as a predictor for the decay of the communities.\\
\textbf{Discussion on \textit{RQ2}}, \textit{What are the network-based properties for the members who left or about to leave community?}: Based on Figure~\ref{fig:performanceResults}, members with less \textit{Betweenness}, less \textit{MinCut}, less \textit{Degree}, less \textit{Closeness}, or less \textit{Coreness} are more susceptible to becoming inactive. This conclusion is also supported by Figure~\ref{fig:correlation_scatter} and by the prediction using machine learning with one attribute and using the STM as shown in Figure~\ref{fig:performanceResults}. The STM can be utilized as a decay indicator when those attributes reach the corresponding $\lambda$ of the members of a community.\\
\textbf{Discussion on \textit{RQ3}}, \textit{How helpful are the exogenous members attributes in predicting members leaving?}: The attributes used, which are based on exogenous information, also showed a potential for providing good prediction results. However, not all of these attributes were helpful. Figure~\ref{fig:featureImportance} suggests that the network-based measures were more important than the exogenous attributes.\\
\textbf{Discussion on \textit{RQ4}}, \textit{Do decayed communities embrace leave patterns that can be used to study the inactivity of communities that are alive?}: Interestingly, the cross-community prediction results shown in Table~\ref{tab:resultsCrossdatasets} suggest that the leave patterns are independent of the community, as we were able to predict the inactivity of a community from the information another one. Apparently, the leave patterns are universal across communities when abstracting the interaction as a network.
\vspace*{-1mm}
\subsection{Threats to validity}
\label{subsec:threatstovalidity}
\textbf{Network Quality:} The networks used in the experiments were constructed from interactions between the members of the StackExchange website, where the nodes of these networks are the members and the edges are the interactions (such as comments) among the members. To guarantee good quality of the network, we took the following steps: For each of the communities we used, a link was considered if it appeared at least one time during the training period. Other values for link persistence overtime yielded sparse network. The training period was selected depending on the number of months a community survived; for example, for the Business Startups community, the training period for constructing the network $G_t$ was $\delta =45$ days. We tried different values for $\delta$. For values $\delta =45 \pm 5$ days, there was no significant difference in the results. For larger values, e.g., $\delta =90$ days, we got few networks that are very dense and can hardly capture any meaningful interaction patters; for smaller values, e.g., $\delta =5$ days, we got too many very-sparse networks. The same argument is applied for the other communities. Thus, we do not expect these design decisions to affect the internal validity of the results.\\
\textbf{Training Quality:} The networks we used are decaying networks, which means that the nearer the network to the time at which community closed, the more inactive member it has. This makes prediction easier for the most recent networks. However, prediction at early time points showed satisfactory results, too, with the ratio between active and inactive members being $55$:$45$. Although we used two different datasets (networks at two different points of time) for training and testing, we used $k$-fold cross-validation, with $k=3$, to further eliminate any random chances of classification bias during the training.
\vspace*{-3mm}
\section{Conclusion and future directions}
\label{sec:conclusion}
\vspace*{-2mm}
Network-based attributes are a good representative of activity behavior in online communities. The STM, which uses only one attribute, is able to effectively predict users' inactivity. The presented method for predicting the decay of the members of online social communities gives information about the attributes of members who became inactive. Those attributes, the network-based attributes and some other community dependent attributes can be used as indicators for the aliveness of an online community. In addition, these attributes can be used to take counter actions when inactivity behavior is detected. Such actions may include, in the context of StackExchange communities, new questions and good answer recommendations as well as additional rewards (like badges and points) for the members. One aspect of the methods contributed in this work is the computational complexity. Some network-based attributes are computationally expensive to compute, especially for large and sparse networks. However, we found that the best results were obtained from attributes that are easy to compute, like \textit{Degree} and \textit{Coreness}~\cite{batagelj2003}. We recommend starting with the STM before using the machine learning classifiers, as the machine learning classifiers are computationally expensive for large datasets. The STM provides good indications regarding which attributes to use. The optimization of the STM is computationally easy for a sorted list; it is $\mathcal{O}(n)$ where $n$ is the number of nodes in the graph. For $2$-month prediction, the upper bounds are $0.85$ and $0.82$ for the F1-score and the accuracy, respectively. For the $4$ months, the upper bounds are $0.91$ and $0.83$ for the F1-score and the accuracy, respectively. The prediction results obtained from nearer time periods to the close time cannot be generalized as the life times of the decayed communities are not equal. Future work will include the following: (1) testing network-based features other than the features presented in this work; (2) testing the method on additional datasets with decay ground truth.
\vspace*{-3mm}
\section*{Acknowledgment}
This research was performed as part of Mohammed Abufouda's Ph.D research supervised by Prof. Katharina A. Zweig.
\vspace*{-2mm}
\bibliography{references}
\end{document}